\documentclass[sigconf,nonacm,balance=false]{acmart}
\usepackage{popets}
\usepackage{multirow}
\usepackage[ruled,vlined]{algorithm2e}
\usepackage{subfigure}

\setcopyright{popets}
\copyrightyear{YYYY}

\acmYear{YYYY}
\acmVolume{YYYY}
\acmNumber{X}
\acmDOI{XXXXXXX.XXXXXXX}
\acmISBN{}
\acmConference{Proceedings on Privacy Enhancing Technologies}
\begin{document}

\title{A Trajectory K-Anonymity Model Based on Point Density and Partition}


\author{Wanshu Yu}
\affiliation{%
  \institution{University of Glasgow}
  \city{Glasgow}
  \country{UK}}
\affiliation{%
	\institution{South China University of Technology, School of Computer Science and Engineering}
	\city{Guangzhou}
	\country{China}}
\email{2817055y@student.gla.ac.uk}
\authornotemark[1]

\author{Haonan Shi}
\affiliation{%
  \institution{Case Western Reserve University}
  \city{Cleveland}
  \state{Ohio}
  \country{USA}}
\affiliation{%
	\institution{South China University of Technology, School of Computer Science and Engineering}
	\city{Guangzhou} 
	\country{China}}
\email{hxs896@case.edu}
\authornote{Both authors contributed equally to this research}

\author{Hongyun Xu}
\affiliation{%
	\institution{South China University of Technology, School of Computer Science and Engineering}
	\city{Guangzhou}
	\country{China}}
\email{hongyun@scut.edu.cn}
\authornote{Corresponding author}


\begin{abstract}
  As people's daily life becomes increasingly inseparable from various mobile electronic devices, relevant service application platforms and network operators can collect numerous individual information easily. When releasing these data for scientific research or commercial purposes, users' privacy will be in danger, especially in the publication of spatiotemporal trajectory datasets. Therefore, to avoid the leakage of users' privacy, it is necessary to anonymize the data before they are released. However, more than simply removing the unique identifiers of individuals is needed to protect the trajectory privacy, because some attackers may infer the identity of users by the connection with other databases. Much work has been devoted to merging multiple trajectories to avoid re-identification, but these solutions always require sacrificing data quality to achieve the anonymity requirement. In order to provide sufficient privacy protection for users' trajectory datasets, this paper develops a study on trajectory privacy against re-identification attacks, proposing a trajectory K-anonymity model based on Point Density and Partition (KPDP). Our approach improves the existing trajectory generalization anonymization techniques regarding trajectory set partition preprocessing and trajectory clustering algorithms. It successfully resists re-identification attacks and reduces the data utility loss of the k-anonymized dataset. A series of experiments on a real-world dataset show that the proposed model has significant advantages in terms of higher data utility and shorter algorithm execution time than other existing techniques.
\end{abstract}

\keywords{Trajectory dataset; Privacy protection; Re-identification attack; Trajectory clustering}

\maketitle

\section{Introduction}
With the rapid development of mobile devices and communication technologies, location-based information online service platforms are widely used, which are highly relevant to people's daily lives and bring convenience. For example, after a user opens a navigation software, the application automatically sends location-based queries to the server, pulling map query results regarding the current area, such as nearby restaurants, car parks, shopping centres, and banks. When a user accesses such a Location-based Information Service (LBS) application, the network operator of the mobile device can extensively record data about their movement trajectory \cite{liu2018location}, i.e. the sequence of location coordinates that the user passed over some time. Releasing the collected information to the public not only facilitates the research work of scientific organizations but also plays a vital role in the transparency of authorities such as operators and governments. However, the publication of data can be exploited by malicious attackers, resulting in the disclosure of user privacy.

Due to the rapid development of high-capacity storage and data analysis techniques, it is possible for attackers to distinguish the trajectory travelled by an individual from publicly released trajectory datasets and to obtain more sensitive privacy information by integrating their location and route with other databases \cite{shaham2020privacy}. Hence it is generally the correspondence between a user's spatiotemporal trajectory and the individual identity. However, it is not sufficient to simply erase the direct unique identifiers from the database to resist attacks. This is because once an attacker combines the quasi-identifiers with known background knowledge, it is possible to deduce the correspondence, thus causing danger to user privacy, property, security and reputation. Therefore, how to scientifically encrypt datasets has become an important issue in data release and privacy protection today.

In order to guarantee the privacy of users despite the public release of trajectory data, it is necessary to employ various techniques to process the trajectory data before releasing it. Many scholars have worked on the issue of trajectory privacy attacks and protection, proposing various techniques to achieve privacy protection in LBS, such as generalization, obfuscation and fuzzing. Nevertheless, although these existing techniques can protect user privacy from being attacked or exposed under certain circumstances, the corresponding algorithms are usually of high time and space complexity. Moreover, due to the specificity of trajectory shape distribution and the sensitivity of location information, the privacy protection processing will lead to the loss of information to a large extent, thus reducing the utility of the data.

To address the problem above, we propose a privacy protection methodology for user trajectories adopting machine learning techniques to prevent revealing the private information of LBS users on the one hand and to retain the features and accuracy of the original trajectories as far as possible on the other hand, so as to reduce the loss of information after data processing. Specifically, it is required that trajectories from different users in the released dataset are indistinguishable from each other. As a result, trajectories in the original dataset typically need to be replaced with the generalized trajectory for several users. The process of thus replacing a specific value with a more general and imprecise value is called generalization \cite{anjum2017banga}. The higher the level of generalization, the higher the extent of privacy protection, but the lower the data utility of the published trajectories and the higher the loss of information after generalization. In order to balance the degree of privacy protection and the generalization information loss, we preprocess trajectories by segmenting them according to the point density and generalize them based on the idea of DBSCAN cluster algorithm \cite{ester1996density}, achieving the resistance of the released dataset to re-identification attacks and preserving the distribution features of the trajectories in the best manner possible. To the best of our knowledge, our paper proposes such a partition preprocessing mechanism for the first time. Our main contributions are summarised as follows.
\begin{itemize}
	\item  We investigate the shortcomings of existing trajectory privacy-preserving algorithms and propose a trajectory K-anonymity model based on Point Density and Partition (KPDP). The deficiencies of the existing models mainly stem from the irregularity of the shape distribution of real trajectories and the specific data structure, making it difficult to measure the similarity between trajectories, and thus unable to accurately cluster and generalize trajectories, resulting in a high information loss in the released dataset relative to the original dataset. Based on this situation, KPDP can segment trajectories based on point density before clustering them so that the length of trajectories is relatively balanced and the spatial distribution characteristics of the original trajectories are retained, yielding a lower generalization information loss than other models.
	
	\item To further enhance the utility of anonymized trajectory datasets and to achieve k-anonymity, this paper proposes an adaptive DBSCAN trajectory clustering algorithm. The algorithm measures the distance between trajectories using the loss from the alignment of trajectories and then clusters them based on sample density. However, due to the uncertainty of the number of samples in the clusters and the possible presence of noise from DBSCAN, direct adoption of its idea cannot guarantee k-anonymity. We consequently developed an adaptive DBSCAN trajectory clustering algorithm that can automatically adjust the values of parameters based on the number of trajectories and noise in each cluster and repeatedly call the core module to cluster. The main advantage of DBSCAN over other unsupervised machine learning-based algorithms is that it is not constrained by given values of parameters and can produce clustering results that better reflect the characteristics of the trajectory distribution, thus improving the data utility of the released dataset.
	
	\item We conducted extensive experiments based on a realistic trajectory dataset to evaluate the privacy-preserving effects of segmentation preprocessing mechanisms and trajectory clustering algorithms under different privacy metrics. The experiment results show that our approach performs better in terms of information loss and running time compared to other existing approaches.
\end{itemize}

The subsequent structure of this paper is organized as follows. Section 2 introduces and defines trajectory privacy attacks, privacy anonymity criteria, privacy-preserving methods, generalization hierarchy models, and trajectory alignment techniques. We then show an overview of KPDP in Section 3. Following this framework, we illustrate the rationale of the segmentation preprocessing mechanism and the design of the anonymization model in Section 4 and 5. The experiment results and evaluation are presented in Section 6. Finally, we conclude with an overview of our contributions in Section 7. 

\section{BACKGROUND AND RELATED WORKS}
\subsection{Attack Model}
A trajectory privacy attack is the acquisition of a user's private information from a trajectory dataset by an attacker with background knowledge. In general, most studies assume that the background knowledge known to the attacker is part of the spatiotemporal points on the user's trajectory, and the privacy information the attacker attempts to disclose is the complete trajectory data of that victim. For a given anonymized dataset, Zhen Tu et al. \cite{8506438} denote the set of users as $U= U_i$ and the corresponding set of trajectories as $T= T_i$, where $T_i$ denotes the spatiotemporal points of the trajectory of user $U_i$. A constant number of partial points sampled from the actual trajectories is considered the attacker's external background knowledge, denoted as $E= E_i$, where $E_i$ denotes the attacker's external observation of the user $U_i$. With any external information $E_i$, an attacker makes a successful re-identification attack if he can match only one trajectory, whose formulation is shown in Eq. (1).
\begin{equation}
	\begin{split}
		C_i=
		\left\{
		\begin{array}{lr}
			1 \qquad \left| {T_j \mid T_j \cap E_j, T_j\in T} \right| = 1, & \\
			&\sum_{i} C_i \geq1 \\
			0\qquad otherwise & 
		\end{array}
		\right.
	\end{split}
\end{equation}

where $C_i$ denotes whether the user $U_i$ is re-identifiable and $\left| \ast \right|$ denotes the size of the set $\ast$.

In addition, adversaries can also launch attacks based on more public information. Zhen Tu et al. \cite{tu2017beyond} stated that an attacker could infer a victim's motivation and behaviour to visit a location by associating the Point of Interest (PoI) that the user passes on a map with the primary function of its corresponding location. Huaxin Li et al. \cite{li2016privacy} matched the locations shared by users on social networks with their real travel trajectories to enable external attackers to infer information such as their age, gender, and education. John Krumm \cite{krumm2007inference} quantified the effectiveness of using different attack algorithms to recognize the location of subjects' homes and then identify them through a programmable web search engine. According to \cite{wernke2014classification}, the types of location privacy attacks explicitly include Single position attack \cite{mokbel2007privacy}, Multiple position attack \cite{beresford2004mix, talukder2010preventing, ghinita2009preventing} and Context linking attack \cite{machanavajjhala2007diversity, gruteser2003anonymous, shokri2011quantifying}. Although there are many approaches to attacking user privacy, re-identification attack remains the most fundamental problem. This paper focuses on studying resistance to privacy issues caused by re-identification attack.

\subsection{Privacy Model}
The protection of individuals from re-identification attacks has been a topic of much discussion in recent years. The k-anonymity criterion is the most commonly used privacy-preserving metric to resist re-identification attacks for data publishing within the privacy and anonymity domain. K-anonymity is a concept introduced by Samarati and Sweeney in 1998. K-anonymity requires that each record stored in a published dataset should be indistinguishable from at least $k-1$ other records \cite{samarati2001protecting, samarati1998protecting, sweeney2002k}, i.e. it requires that the same quasi-identifier refers to at least multiple records, making it impossible for adversaries to connect records with other databases by quasi-identifiers and thus deduce user identity and more private information.
Current k-anonymity implementations are mostly used to protect data anonymity for category and numerical attributes in general relational databases, including Generalization and suppression, Incognito, Top-down specialization, Clustering, and Multidimensional partitioning \cite{fung2005top, lefevre2005incognito, lefevre2006mondrian}. However, for such irregular geometric data structure as trajectory, it requires a specific processing method to achieve k-anonymity \cite{bettini2009anonymity, gruteser2003anonymous, xu2008exploring, shokri2010unified}.

In this paper, we need to ensure at least $k$ distinct trajectories in each cluster obtained from the original trajectory set and generalize them to identical anonymous records to form a trajectory dataset that conforms to k-anonymity.

\subsection{Defense Techniques}
Among diverse researches to achieve k-anonymity of trajectory data, generalization is one of the most dominant approaches. According to the different details of generalization techniques, such as the encoding and operation of the Domain Generalization Hierarchy (DGH) tree, there are three main types of generalization: full domain generalization, subtree generalization and cell level generalization \cite{yaseen2018improved}. Acar Tamersoy et al. \cite{tamersoy2012anonymization} proposed a heuristic approach based on the concept of generalization to achieve k-anonymity. Sina Shaham et al. \cite{shaham2020privacy} used a heuristic and a variant k-means algorithm for trajectory clustering and anonymization. Marco Gramaglia et al. \cite{DBLP:journals/corr/GramagliaFTB17} used a k-normalization algorithm to address the efficiency problem of generalization during the anonymization of trajectory datasets. 

In addition to generalization methods, many researchers have worked on resisting trajectory privacy attacks from multiple other perspectives. The authors of \cite{ghinita2009preventing, cicek2014ensuring, naghizade2014protection} provide special treatments for sensitive locations on maps to protect the semantic privacy of trajectories. Zhen Tu et al. \cite{8506438} protect trajectories from re-identification and semantic attacks based on k-anonymity, l-diversity and t-confidentiality. There are also researches which generate stopping points and noise points to obfuscate the original trajectory set, demonstrating the effectiveness of resistance to virtual locationinformation \cite{chow2009faking, jiang2013publishing, do2016another}. Jiaxin Ding \cite{ding2015trajectory} prevents an attacker from identifying a specific user's trajectory by exchanging the user's ID at the intersection of the trajectory. Jae-Gil Lee et al. \cite{lee2007trajectory} sliced the trajectory into line segments and clustered the new set of segments based on a definition of the distance between the segments. A middleware structure and an algorithm for adjusting the resolution of location information along the spatial or temporal dimension was introduced by \cite{gruteser2003anonymous}, which satisfies a specified anonymity constraint within a given region. 

This paper proposes a trajectory k-anonymity approach to preserve privacy via generalization techniques with low loss of data utility and algorithm time complexity. 

\subsection{Generalization for Secure Data}
Generalization techniques enable the goal of ensuring the privacy of published dataset without compromising data availability. Generalization and suppression\cite{sweeney2002k} are used to provide privacy to individuals. Pre-defined generalization hierarchy\cite{iyengar2002transforming} allows the construction of generalization hierarchies before data masking. Full domain generalization hierarchy\cite{samarati2001protecting} enables the mapping of attributes to a more general domain in the domain generalization hierarchy.

A DGH tree is a quantitative model of information loss for generalizing numerical or categorical attributes \cite{jiang2013publishing}. In the anonymity domain, the structure of DGH trees has several different ways of formation. The structure of a DGH tree for numerical attributes can either be predefined by the user based on the usage scenario \cite{iyengar2002transforming} or built dynamically during the generalization process \cite{campan2011fly}. Category attributes often require the manual creation of hierarchical structures considering attribute characteristics and usage scenarios \cite{sweeney2002achieving}. Due to the complexity of the practical situation, not all anonymization processes for category attributes apply to the DGH tree model. For the generalization of the trajectory dataset in this paper, the latitude and longitude of the trajectories are usually used to dynamically construct the DGH tree and generalize the trajectory set based on this model while calculating the caused information loss.

\subsection{Trajectory Alignment}
Dynamic Sequence Alignment (DSA) is a trajectory alignment algorithm derived from the sequence alignment method of proteins and DNA in biology \cite{chen2017cmsa, le2017protein}. The algorithm uses a dynamic programming approach to obtain the loss matrix for the alignment of two trajectories recursively, and then from the backtracking of this loss matrix, find the strategy that can make the merging of these two trajectories produce the least information loss and subsequently obtain the merged trajectory and the minimum information loss value.

Based on this, the Progressive Sequence Alignment (PSA) algorithm is derived as a multi-sequence alignment algorithm capable of aligning a cluster of trajectories \cite{chowdhury2017review}. The PSA algorithm can sequentially perform DSA operations on the trajectories within a cluster to obtain a synthetic trajectory obtained by aligning all trajectories within the cluster, and the PSA algorithm is often used to generalize the clusters of trajectories formed by clustering because it first selects the longest trajectory from a group of trajectories as the base trajectory, and then sequentially selects the remaining trajectories within the group in order of trajectory length from longest to shortest to align and synthesize with the base trajectory based on DSA. The generalized trajectory generated after DSA process will become the new base trajectory for the subsequent DSA process until all trajectories in the group have been aligned with the base trajectory.

\section{SYSTEM OVERVIEW}
\subsection{System Utility Measurement}
In the KPDP framework, trajectory alignment is the key to performing trajectory anonymization, and information loss is incurred in trajectory alignment. In order to calculate the loss of KPDP in the process of anonymizing trajectories more accurately and efficiently, a new DGH tree is proposed in this paper. This DGH tree is a partially ordered tree structure, which is able to map the specific and generalized values of attribute $A$ for a certain attribute $A$. The root node of the DGH tree indicates the case with the highest degree of generalization. Our DGH tree is constructed by dividing a number of small intervals of equal length within the range of corresponding values taken and then using these small intervals as leaf nodes to construct a full binary tree. If the number of leaf nodes is not enough to fill the bottom level of the binary tree, some invalid points are added to fill it up. A simple illustration of a DGH tree with a 4-layer structure is shown in Figure 1. The leaf nodes numbered 12 can be generalized to the parent node 6 or the ancestor nodes 3 and 1. Specifically, the DGH trees of KPDP in this paper are two DGH trees formed by building latitude and longitude in the trajectory set, corresponding to the x-axis and y-axis coordinate systems on the map plane space, respectively.
\begin{figure}[h]
	\centering
	\includegraphics[width=\linewidth]{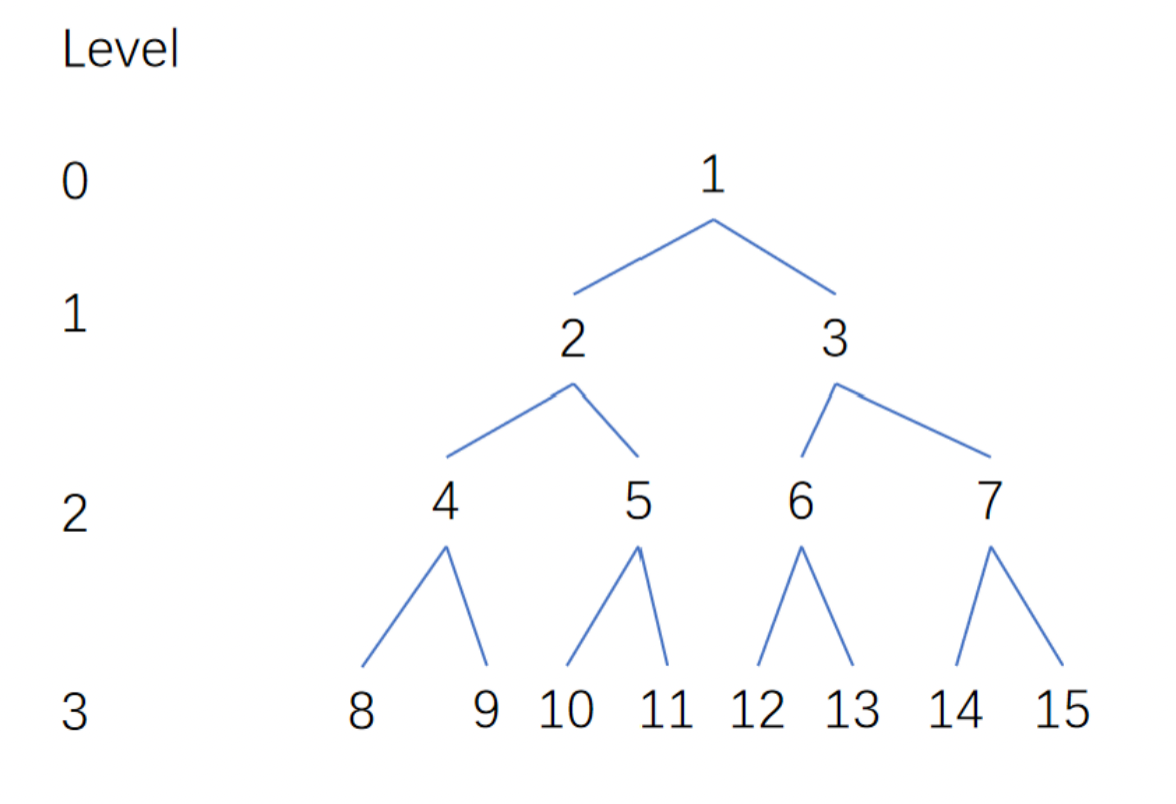}
	\caption{Schematic diagram of the DGH tree structure used in the utility measurement of KPDP}
	\Description{A full binary tree with nodes 8, 9, 10, 11, 12, 13, 14, 15 at the bottom level. 4, 5, 6, 7 at the penultimate level. 2, 3 at the penultimate level. 1 at the top level.}
\end{figure}

For KPDP, the information loss generated by this system mainly comes from the generalized information loss in the process of satisfying the k-anonymity criterion. The calculation of generalized information loss is based on the relationship between nodes on the DGH tree. The generalized information loss includes single-node generalized information loss as well as multi-node generalized information loss.

\textbf{Definition 1. Single-node generalized information loss:} The information loss incurred when generalizing a node to a parent or higher level node is calculated as shown in Eq. (2).
\begin{equation}
	Loss_g(node_i, node_j)=log_2(LF(node_i))-log_2(LF(node_j))
\end{equation}

Where $Loss_g(node_i, node_j)$ is the generalization information loss generated by generalizing $node_j$ to $node_i$,  $LF(node_k)$returns the number of leaf nodes owned by $node_k$. The special case of leaf nodes being generalized to the root is called suppression\cite{shaham2020privacy}, and in the suppression case, the generalization information loss is calculated as shown in Eq. (3). 
\begin{equation}
	Loss_g(node_i)=H
\end{equation}
Where H denotes the height of the DGH tree.

\textbf{Definition 2. Multi-node generalized information loss:} Any two nodes on the DGH tree need to be generalized by finding the smallest subtree containing both nodes. The Lowest Cmmon Ancestor (LCA) of two nodes is the result of their generalization. The information loss caused by generalizing two nodes to their LCA nodes is calculated as shown in Eq. (4).
\begin{equation}
\begin{aligned}
	Loss_g(node_i,node_j,node_{LCA})=Loss_g(node_{LCA},node_i)\\
	+Loss_g(node_{LCA},node_j)
\end{aligned}
\end{equation}

Since the trajectories input to KPDP system usually has irregular geometry, in order to cluster different trajectories to achieve the purpose of trajectory anonymization of KPDP, this paper uses PSA algorithm in order to cluster multiple trajectories in PSA. We need to calculate the trajectories with the smallest relative distance and the closest shape for clustering to achieve the purpose of trajectory anonymization. In order to complete the calculation process of PSA, this paper adopts the DSA algorithm to calculate the distance between trajectories and the information loss generated in the process of clustering trajectories and uses the information loss generated by trajectory alignment as a measure of the relative distance between trajectories in clustering. In DSA, the generalization information loss of generalizing two trajectory points and suppressing a certain trajectory point is calculated based on the DGH tree generalization model with the corresponding dimensional attributes. According to Eq. (3) and Eq. (4), for any two trajectories and, when DSA is performed on these two trajectories, the recursive equation of dynamic programming is shown in Eq. (5).
\begin{equation}
\begin{split}
		&SAmatrix[i][j]=\\
		&min\left\{
		\begin{array}{lr}
		SAmatrix[i-1][j-1]+(Loss_g(p_i.X, q_j.X, X_{LCA})\\
		+Loss_g(p_i.Y,q_j.Y,Y_{LCA})), \\
		SAmatrix[i][j-1]+(Loss_g(q_j.X)+Loss_g(q_j.Y)),\\
		SAmatrix[i-1][j]+(Loss_g(p_i.X)+Loss_g(p_i.Y)) 
		\end{array}
		\right.
\end{split}
\end{equation}

\subsection{KPDP Workflow}
KPDP is mainly composed of two parts, which are the Partition model and the Anonymization model, the trajectory dataset of multiple users is the input of KPDP, and the anonymized trajectory dataset is the output of KPDP. In this case, because the length difference of two trajectories close to each other is large, the information loss from DSA alignment is large, and thus the two trajectories cannot be grouped into one cluster in the clustering algorithm based on the distance of the trajectories, thus makes the clustering in Anonymization model less effective and generates a larger information loss. As shown in Figure 2, it can be found from Eq. (4) that in the process of aligning trajectory $tr_1$ with trajectory $tr_2$, $p_1$ and $q_1$, $p_2$ and $q_2$ are generalized to multiple nodes, while $q_3$, $q_4$, $q_5$ are generalized to the root node of DGH tree by a single node, and this process will produce excessive information loss. In this paper, we set up a Partition model to reduce the information loss of KPDP anonymization while ensuring the requirement of KPDP anonymization. 
\begin{figure}[h]
	\centering
	\includegraphics[width=\linewidth]{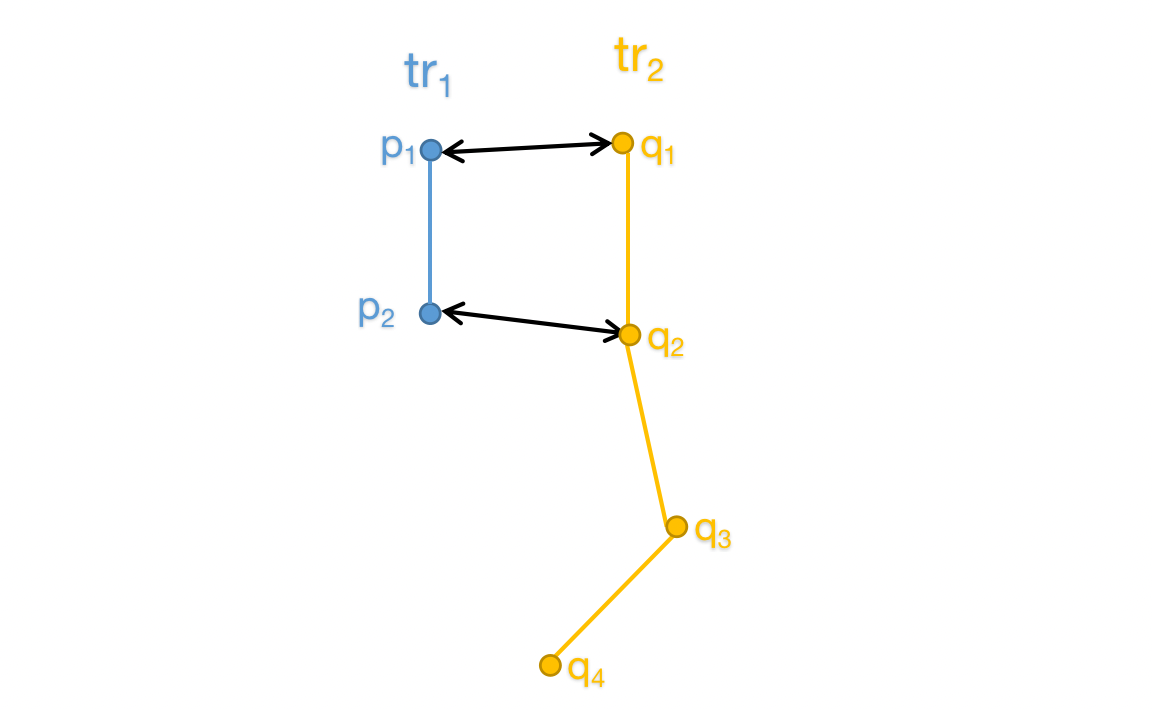}
	\caption{Schematic diagram of the DGH tree structure used in the utility measurement of KPDP}
	\Description{Blue trajectory tr1 and yellow trajectory tr2 are aligned, p1 with black arrow pointing to q1 and p2 with black arrow pointing to q2}
\end{figure}

The specific workflow is shown in Figure 3. The trajectory dataset needs to be preprocessed by the Partition model first, which enables all trajectories to be processed in advance to keep the original geometric features of trajectories in the Anonymization model as much as possible, as well as to prevent excessive information loss in the process of Anonymization model. The partition model prevents the loss of information in the process of the Anonymization model. The processed datasets are transferred from the Partition model to the Anonymization model, which uses the PSA algorithm and the adaptive DBSCAN clustering algorithm proposed in this paper to complete the trajectory clustering, and finally outputs the anonymized trajectory datasets of the KPDP system.
\begin{figure}[h]
	\centering
	\includegraphics[width=\linewidth]{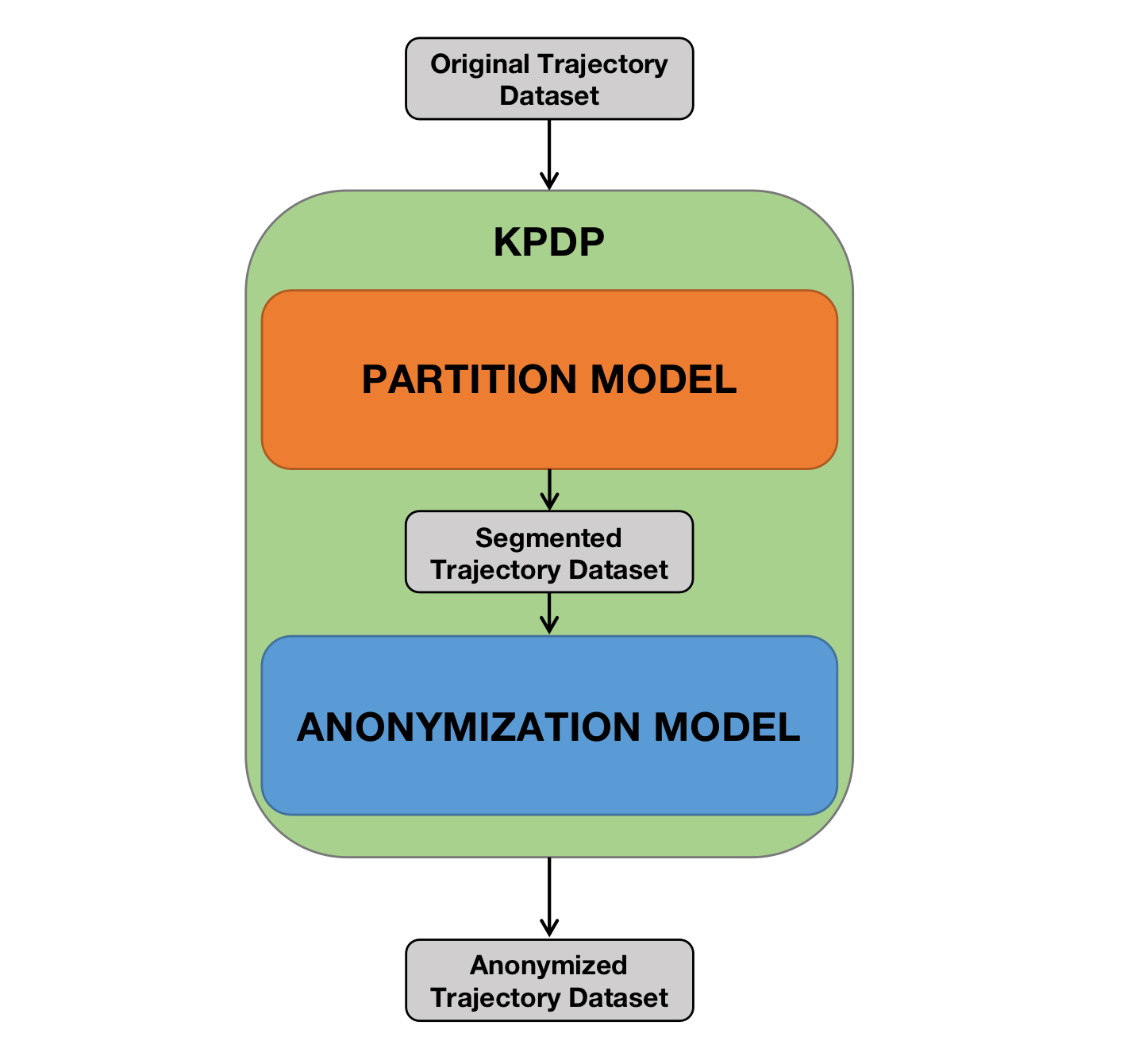}
	\caption{KPDP Workflow}
	\Description{The original trajectory dataset is pointed to the KPDP with an arrow, which contains a partitioned model and an anonymized model in the KPDP, the partitioned model is pointed to the anonymized model with an arrow, and the KPDP output is pointed to the anonymized trajectory dataset with an arrow.}
\end{figure}

\section{PARTITION MODEL}
Based on the workflow of KPDP in Section 3.2, this section focuses on the segmentation preprocessing of trajectories to reduce the generalization information loss of trajectories afterwards. This process refers to segmenting the trajectories based on point density before anonymizing the trajectory set so that the released dataset will retain the distribution features of the trajectories and reduce the generalization information loss in the alignment and clustering steps as much as possible. We illustrate the three main steps of the partition model - generating auxiliary points on the trajectory, then clustering the point set, and segmenting the trajectory based on the clustering distribution. The steps are interlocked to make the length of trajectories relatively average. The specific segmentation preprocess of the original trajectory set is schematically shown in Figure 4. 
\begin{figure}[h]
	\centering
	\includegraphics[width=\linewidth]{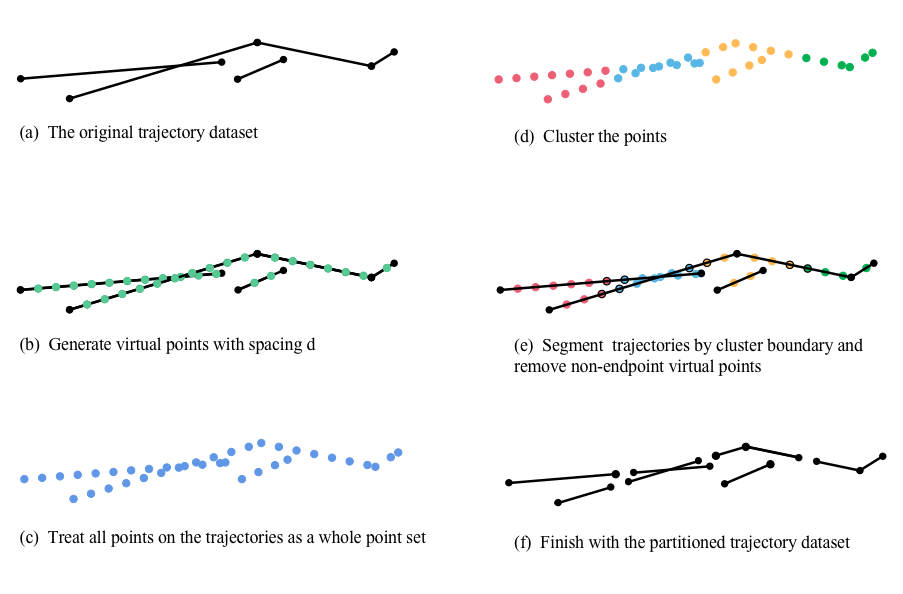}
	\caption{Schematic diagram of the preprocessing}
	\Description{Blue trajectory tr1 and yellow trajectory tr2 are aligned, p1 with black arrow pointing to q1 and p2 with black arrow pointing to q2}
\end{figure}
First, for the original trajectory set in Figure 4(a), Figure 4(b) shows the generation of green auxiliary points on the trajectory with the same distance $d$. All the points in Figure 4(c), including the actual existing and virtual auxiliary points, are considered whole point sets for clustering. The clusters of points in Figure 4(d) are distinguished from each other by different colours, i.e., points of different colours belong to different clusters. These points are mapped back to the trajectory set in Figure 4(e), and the trajectories are partitioned at the neighbouring points belonging to different clusters according to the boundaries of the point clusters. The final result is shown in Figure 4(f), where the auxiliary points as trajectory endpoints after segmenting are kept in the trajectory set and form a new segmented dataset with other actual points.

We conducted extensive experiments to evaluate our segmentation model. The results demonstrate that, compared with the direct method of clustering and generalizing the trajectories, adding the preprocessing step can not only effectively reduce the overall generalization information loss but also speed up the running of the trajectory clustering algorithm.

\subsection{Auxiliary Point Generation}
Before segmenting the trajectory, relatively dense auxiliary points are added between adjacent points. These virtual auxiliary points are equally spaced, as shown in Figure 4(b). Since the actual points on the trajectory are time-ordered, the primary purpose of setting auxiliary points is to make line segments of different lengths have the same effect on the density of points in their neighbourhoods so that the line segments represented by points can be more similar to the solid form of the line in space. auxiliary points are defined as follows.

\textbf{Definition 3. Auxiliary Point: }Points that do not exist in the trajectory dataset and are used to reflect the spatial distribution structure of the trajectory. For the line segment formed between two time sequence adjacent points, starting from the end of the previous time sequence, a auxiliary point is added for each fixed distance $d$ along the line segment.

The smaller the distance between the generated auxiliary points, the better the point set can reflect the distribution shape of trajectories in space. In contrast, if the spacing is too small, it will increase the amount of processed data and affect the operation efficiency. 

\subsection{Point Set Clustering}
In order to make the length of trajectories relatively uniform, partitioning trajectories based on the difference in spatial trajectory density is our proposed solution. Regarding distribution, the density of trajectories in macroscopic space is reflected as the density difference of the points on the microscopic level. The point clustering algorithm can automatically gather the close points into clusters, reflecting the density distribution of points on the plane space.

For the trajectory dataset with auxiliary points, all coordinate points on the trajectory are regarded as the whole point set to be clustered. Meanwhile, the mapping relationship between each point and the cluster it belongs to is recorded for the subsequent segmentation operation.

We called the k-means point clustering method from the machine learning Sklearn library to divide the points into $k$ clusters based on the spatial Euclidean distance between them. The k-means algorithm is one of the most basic and widely used clustering algorithms that can divide data samples with different attribute values into a designated number of clusters and use the mean of all samples within each cluster as the representative points \cite{macqueen1967classification}. The main idea is to divide the data set into different classes by iteratively adjusting the clustering centres so that the mean error criterion function, which measures the clustering performance, is optimal, thus ensuring that the generated clustering results are compact within clusters and sparse from each other.

The effective operation of the clustering algorithm is generally based on the homogenization and standardization of the data feature variables. Since the attribute values used to calculate the Euclidean distance between trajectories only contain two dimensions, longitude and latitude, and there is no significant data disparity with uniform magnitude, the k-means algorithm can be directly applied to divide the point set on the plane space map into clusters.

\subsection{Trajectory Segmentation}
In this stage, we use the clustering boundaries generated by the k-means algorithm to segment the trajectories to reduce the disparity in the length of trajectories in the original dataset. Referring to \cite{lee2007trajectory}, the sum of segmented trajectories is not necessarily the original trajectory but a characteristic reflection of its structure distribution. Therefore, when trajectory clustering is performed later, the segments of a trajectory may belong to several different clusters and subsequently be generalized to different anonymous trajectories. However, the accuracy of the trajectory clustering will be relatively higher due to the reduced cost of information loss when aligning long and short trajectories later. In contrast, the overall trajectory clustering will lose more detailed features and incur higher generalization information loss during generalization. In the KPDP framework, after clustering the segmented trajectories, the length of trajectories within each cluster is relatively consistent, so the shape of the anonymous trajectories will be more reasonable.

The segmentation process of the trajectory set is described as follows. Iterate through each trajectory in the trajectory dataset containing auxiliary points and check whether the adjacent points on a trajectory belong to the same cluster. If they are not the same, the trajectory is segmented, and a new trajectory is generated. When the endpoint of a segmented trajectory is a auxiliary point, it will be added to the newly generated trajectory dataset as a real trajectory point, while other non-endpoint auxiliary points will be removed and will not be involved in the subsequent privacy-preserving processing. The pseudo-code for generating a segmented trajectory dataset is shown in Algorithm 1, whose input is the trajectory dataset containing auxiliary points.

\begin{algorithm}
	\SetKwData{Left}{left}
	\SetKwData{This}{this}
	\SetKwData{Up}{up} 
	\SetKwFunction{Union}{Union}
	\SetKwFunction{FindCompress}{FindCompress} 
	\SetKwInOut{Input}{input}
	\SetKwInOut{Output}{output} 
	
	\Input{Dataset $T$} 
	\Output{Partitioned Dataset $T_{partitioned}$} 
	\BlankLine 
	
	\emph{Let $T_{partitioned}$ be an empty set that will store the new partitioned trajectory dataset}\; 
	\For{$tr$ \textbf{in} $T$}
	{ 
		\emph{Let $new_{tr}$ be an empty set that will store the new trajectory}\;
		\emph{Append tr[0] as the first point to $new_{tr}$}\;
		\For{$p$ \textbf{in} $range(0,len(tr)-1)$}
		{\eIf{the point p and the adjacent point p+1 belong to the same cluster}{Append tr[p+1] to $new_{tr}$}
		{\If{point p is not a real point}{Append tr[p] to $new_{tr}$}
		\emph{Append $new_{tr}$ to $T_{partitioned}$}\;
		\emph{Let $new_{tr}$ be an empty set that will store the new trajectory again}\;
		\emph{Append tr[p+1] to $new_{tr}$};
		}}
		\emph{Append $new_{tr}$ to $T_{partitioned}$}\;
		{\label{forins} 
		  \caption{Trajectory segmentation algorithm}
		  \label{algo_disjdecomp} 
		} 
	}
	\textbf{return} $T_{partitioned}$
\end{algorithm}

On the one hand, the maximum value of distance lost in segmenting the trajectory is $d$ because the spacing will not be smaller than the distance between the adjacent auxiliary points and the actual point or between two actual points when generating virtual auxiliary points along the trajectory direction before. In order to make the segmented trajectory closer to the original one, the parameter $d$ should be as small as possible without causing the algorithm to be overly complicated so that the loss due to segmentation can be minimized when cutting the line segment between two adjacent points. On the other hand, the trajectory segmentation should not only ensure accuracy but also have simplicity, i.e., use as few points as possible to characterize the shape of the trajectory. The virtual auxiliary points that are not endpoints on the trajectory do not contribute significantly to the subsequent generalization process of the trajectory but rather increase the time complexity of the alignment algorithm, so they are discarded when generating the new segmented trajectory dataset.

\section{ANONYMIZATION MODEL}
In order to achieve the anonymity requirement, we introduce clustering algorithms that can gather data samples based on similarity. Clustering is an unsupervised learning method in the field of machine learning that is capable of discovering patterns implicit in a dataset. By clustering the preprocessed trajectory set wisely, it can produce a low information loss during generalization and anonymization, further maintaining the distribution characteristics and data utility of the original trajectory set. In this paper, two trajectory clustering algorithms are considered to construct the anonymization model, respectively, the iterative k’-means algorithm and the adaptive DBSCAN algorithm. Both use the alignment information loss obtained by DSA as the distance indicator between two trajectories and provide a design such that the number of trajectories contained in each cluster is no less than $k$, ensuring compliance with the privacy-preserving requirement of k-anonymity. Among them, the adaptive DBSCAN algorithm is the primary one that this paper focuses on as a method that can significantly improve the utility of the anonymous trajectory dataset and reduce the model running time, while the iterative k’-means algorithm is mainly used for comparison. These two algorithms run independently in the anonymization model of KPDP. After clustering, KPDP will apply PSA to generalize the trajectories of each cluster to derive the anonymous trajectory set for publication.

\subsection{Iterative K'-means algorithm}
We borrowed the idea from \cite{shaham2020privacy} to perform k’-means clustering on trajectories (where the “’” is used to distinguish the “k” that has different meanings in k-means and k-anonymity) and ensure the number of trajectories within each cluster is at least $k$ by iteration. K’-means is a distance-based clustering algorithm. Its clustering similarity is calculated using the mean distance between objects within each cluster. The brief idea is to divide data objects into $k’$ clusters according to the input value of $k’$, making the similarity within each cluster higher and the similarity between different clusters lower. The iterative k’-means algorithm is used for comparison with the adaptive DBSCAN algorithm.

The basic k’-means algorithm works by first selecting any $k’$ objects from the dataset as the initial cluster centers and assigning the remaining objects to the most similar clusters (i.e., closest in the distance) to them based on their similarity (usually Euclidean distance). Then for each cluster, a new cluster center is calculated based on the mean value of the distances of all objects in the cluster. This process is repeated until the cluster centers no longer change or the standard measure of clustering performance converges.

Measuring the relative distance between trajectories is a major difficulty with an irregular data structure like spatiotemporal trajectories. The iterative k’-means algorithm in this paper uses the information loss generated by DSA of two trajectories to measure the relative distance of trajectories. In addition, when designing the trajectory clustering algorithm based on k’-means, many technical details need to be adjusted according to the characteristics of the trajectory data and the rationality of the processing method so that the iterative k’-means algorithm can effectively cluster and generalize the trajectories in the trajectory privacy protection model. Its workflow is described as follows: \textbf{(1)} Calculate the initial number of clusters based on the value of $k$ required for k-anonymity and the number of trajectories in the dataset. \textbf{(2)} A randomly selected trajectory from the trajectory set is used as the initial clustering center for each cluster. \textbf{(3)} Assigning all trajectories in the trajectory set to the cluster center that produces the least loss of alignment information with its DSA. \textbf{(4)} Apply the PSA algorithm to each cluster and generalize and merge the trajectories it contains to form a new cluster center. \textbf{(5)} Repeat steps (3) and (4) until the trajectories contained in each cluster no longer change, completing k’-means clustering of trajectories.
\textbf{(6)} Dissolve the clusters containing less than $k$ trajectories and repeat the steps of k’-means clustering until all clusters conform to k-anonymity.

Compared with the basic k'-means algorithm, the iterative k'-means algorithm gets the centers of a cluster of trajectories by PSA, except for the initial clustering centers randomly selected from the set of trajectories. In addition, the ordinary k'-means algorithm determines whether to perform the next clustering iteration based on the change of the cluster centers. However, due to the specificity of the generalization trajectory, when the cluster assignment is no longer changed, it marks the iteration stop to reduce the algorithm complexity.

Theoretically, the iterative k'-means algorithm is random for selecting initial clustering centers. This may lead to a high overall generalization information loss by generalizing each cluster when the distribution of initial clustering centers is poor, reducing the data utility of the final trajectory set used for publication. The experimental performance of the iterative k'-means algorithm on real datasets will be discussed in Section 6.

\subsection{Adaptive DBSCAN Algorithm}
Inspired by the iterative k'-means algorithm, we propose the adaptive DBSCAN algorithm, which can capture the distribution characteristics among trajectories with more details. DBSCAN is a density-based spatial clustering of applications with noise, which measures the similarity between data samples in terms of density \cite{ester1996density}. Compared with k'-means, DBSCAN does not need cluster centres to instruct clustering. Instead, it searches for high-density regions separated by low-density regions through the density connectivity of the samples. These separated high-density regions are the corresponding clusters to which the corresponding samples belong. In contrast to k'-means, which is unable to discover spherical clusters, DBSCAN can not only discover clusters of arbitrary shapes but also has special treatment for noisy samples to suppress the influence of abnormal data on clustering.

The basic DBSCAN algorithm requires two parameters to be entered before working: the neighbourhood radius $epsilon$ and the minimum number of samples contained in the neighbourhood $minPts$. Once it starts running, the DBSCAN algorithm will traverse and label each sample in the dataset. First, for any sample that has not been labelled, find all samples whose relative distance to it is within $epsilon$. If the number of samples contained in the neighbourhood of the sample reaches the threshold indicator $minPts$, the sample and all samples in its neighbourhood will form a cluster, and the sample will be marked as visited. Then recursive processing is performed for the other samples in that cluster to extend the cluster by the same steps.

Conversely, if the number of samples contained in the neighbourhood of that sample is less than $minPts$, the sample is temporarily marked as noise. Once the recursion is over, the cluster has been sufficiently extended, i.e. all samples are marked as visited. The algorithm then proceeds to traverse the points in the dataset, and the points that have not been labelled are processed similarly. The basic DBSCAN algorithm outputs clusters from sample density expansion and possibly noisy samples that are still labelled as the noise at the end of the algorithm.

We conducted an intensive study on the utilization of DBSCAN ideas for the trajectory clustering algorithm and proposed an adaptive DBSCAN algorithm that meets the privacy preservation requirement. Similar to the iterative k'-means algorithm, the adaptive DBSCAN algorithm measures the relative distance of two trajectories by the information loss generated by DSA. In order to make the anonymous trajectory dataset generated by clustering and generalization fulfil the k-anonymity criterion, we assign $k$ as the value of $minPts$ in the adaptive DBSCAN algorithm. This is because the parameter $minPts$ is the threshold indicator of whether a trajectory is clustered with its neighbouring trajectories, so as long as the value of $minPts$ is greater than $k$, it can ensure that the number of trajectories within each cluster is at least $k$, resulting in k-anonymity of the trajectory dataset.

As for the noisy samples that may exist in DBSCAN, we also handled them specifically in the trajectory privacy-preserving scenario. Analogous to the iterative k'-means algorithm, the adaptive DBSCAN algorithm repeatedly calls the core DBSCAN code until all clusters satisfy k-anonymity in order to make the noisy trajectories eventually satisfy the anonymity requirement as well. The noisy trajectories formed by DBSCAN each time will become the new input dataset for the next clustering, while another input parameter, the neighbourhood radius $epsilon$, will be enlarged appropriately to lower the judgment criterion of density connection between samples so that those noisy trajectories can be clustered more easily. The algorithm will not stop calling the DBSCAN core code until there are no more noisy samples in the dataset.

The pseudo-code of the adaptive DBSCAN algorithm is shown in Algorithm 2. It takes the trajectory dataset, the value of $k$ of the k-anonymity criterion and the neighbourhood radius parameter $epsilon$ as inputs, and it outputs the clustered trajectory dataset. Its workflow is described as follows: \textbf{(1)} For trajectories that have not been labelled in the trajectory set, \textbf{(2)} if the number of trajectories with the information loss generated by DSA with that trajectory is less than the neighbourhood radius $epsilon$ is greater than the threshold $minPts$ (which takes the value of $k$), find all trajectories connected to that trajectory to form a cluster, and mark all trajectories in the cluster. \textbf{(3)} otherwise, mark the trajectory as noise, find the next unmarked trajectory, and repeat the previous step until all trajectories are marked. \textbf{(4)} For the set of trajectories that are still marked as noisy at this time, adaptively enlarge the value of the neighbourhood radius epsilon and repeat the above steps until all the generated clusters satisfy k-anonymity.
\begin{algorithm}
	\SetKwData{Left}{left}
	\SetKwData{This}{this}
	\SetKwData{Up}{up} 
	\SetKwFunction{Union}{Union}
	\SetKwFunction{FindCompress}{FindCompress} 
	\SetKwInOut{Input}{input}
	\SetKwInOut{Output}{output} 
	
	\Input{Dataset $T$, Anonymity Criterion$k$, Neighbor Radius $epsilon$} 
	\Output{Trajectory Cluster Dataset $T_{clus_{k}}$} 
	\BlankLine 
	
	\emph{Let $T_{clus_{k}}$ be an empty set that will store the clusters with at least $k$ trajectories}\; 
	\While{true}{
		$T,T_{clus_{k} } \leftarrow    \textbf{TrajectoryDBSCANClustering}(T, epsilon, k)$\;
		\emph{Append the cluster in $T_{clus}$ to $T_{clus_k}$}\;
		\If{$\left| T \right| < 2 * k$}{
			\emph{Cluster $T$'s remaining trajectories together and append the last cluster to $T_{clus_k}$}\;
			\textbf{break}
		}
		\eIf{$epsilon < top_{epsilon}$}{Increase the value of $epsilon adaptively$}{Cluster $T$'s remaining trajectories together and append the last cluster to $T_{clus_k}$\; 	\textbf{break}}
	}
	{\label{forins} 
		\caption{Adaptive DBSCAN algorithm}
		\label{algo_disjdecomp} 
	} 
	\textbf{return} $T_{clus_{k}}$
\end{algorithm}

In the loop of the algorithm, the value of neighbourhood radius $epsilon$ will be changed adaptively based on the statistical distribution of the relative distance between trajectories. For example, in the first several rounds, the density of the trajectories is high, and the relative distances between trajectories will be concentrated at a low level. If we increase the neighbourhood radius $epsilon$ by a small margin, we can efficiently cluster the trajectories in the area of high density. As looping times increase, the number of trajectories with low relative distances decreases, and the main distribution of distances among trajectories will tend to a higher value range. At this point, the neighbourhood radius $epsilon$ must be raised by a greater magnitude to reduce invalid loops of the core function and improve the efficiency of the adaptive DBSCAN algorithm.

Theoretically, the time complexity of the adaptive DBSCAN algorithm will be an order of magnitude lower than the iterative k'-means algorithm due to no iteration of cluster centres required, which significantly reduces the time consumption in the trajectory clustering session. In addition, the algorithm can specialize noisy data to enhance the data utility of the generalized trajectories, and the stability of the trajectory cluster generation process is also an advantage it has. As for the logic of the algorithm, how to adaptively adjust the value of the neighbourhood radius parameter is the key point to improve the operation efficiency. Reducing the algorithm’s complexity and optimizing the parameter values are unavoidable contradictions requiring a balanced algorithm design. Extensive experiments on a real dataset will evaluate the performance of our proposed model.

\section{EVALUATION}
This section describes the experiments on trajectory privacy protection of the KPDP framework against re-identification attacks on real-world datasets. We mainly evaluate and analyze the effectiveness of the preprocessing step of the trajectory set and the performance of two trajectory clustering algorithms and explore the role of different values of parameters in the k-anonymity criterion on the experimental procedure and results. The experimental results reflect the superior performance of our method in all aspects.
\subsection{Dataset Introduction}
The trajectory dataset used in the experiment is from the Geolife project \cite{zheng2010geolife, zheng2009mining, zheng2008understanding} and the T-Drive dataset \cite{yuan2010t, yuan2011driving}, which consists of GPS trajectories of mobile device users in the Beijing area, specifically including the longitude and latitude information and time series relationship of trajectory points. After obtaining the basic trajectory data, the original trajectory set used for trajectory privacy protection in this paper is all the trajectories intercepted in an area on the map of Beijing, China, corresponding to the latitude and longitude ranges of $116.300000\sim116.316000^\circ E$ and $39.989500\sim40.000000^\circ N$. The road network model composed of this trajectory set is shown in Figure 5, where The trajectory consists of longitude and latitude coordinate points collected after a certain time interval.

\begin{figure}[h]
\centering
\includegraphics[width=\linewidth]{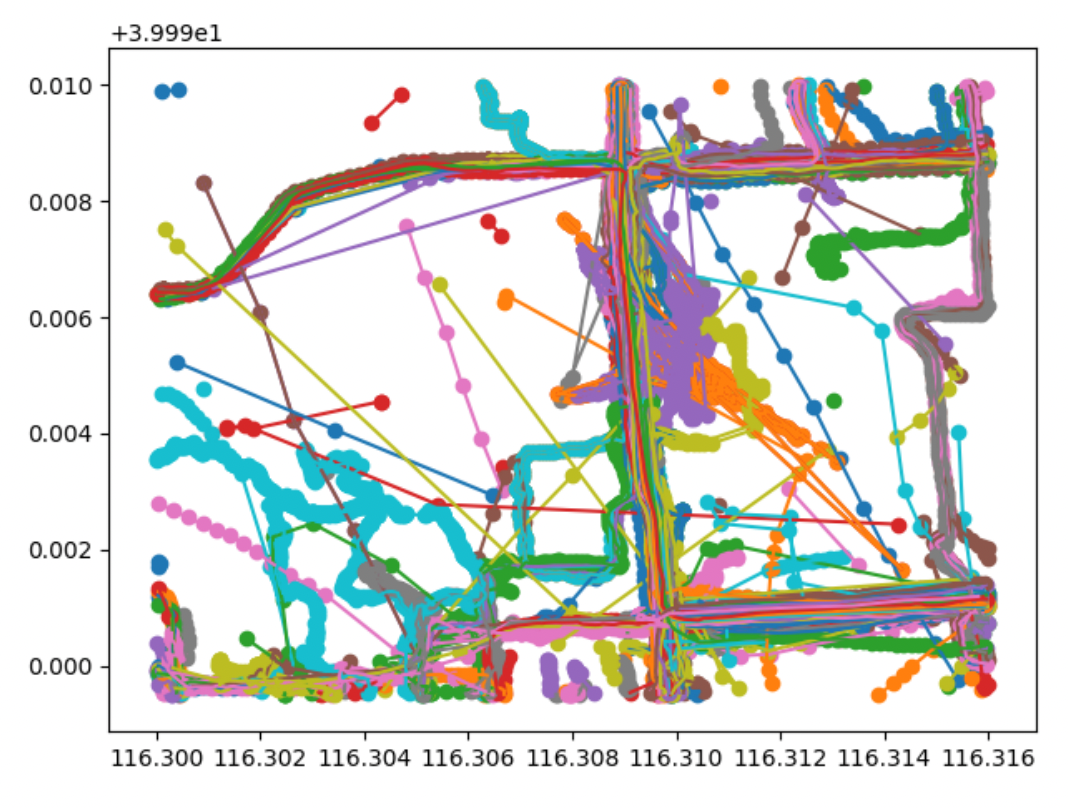}
\caption{Road network mapping of the original trajectory set}
\Description{Trajectory set road network diagram, different trajectories are marked with different colors, trajectory points are represented by circles, and trajectory points are connected to trajectory points by lines}
\end{figure}

\begin{figure}[h]
	\centering
	\includegraphics[width=\linewidth]{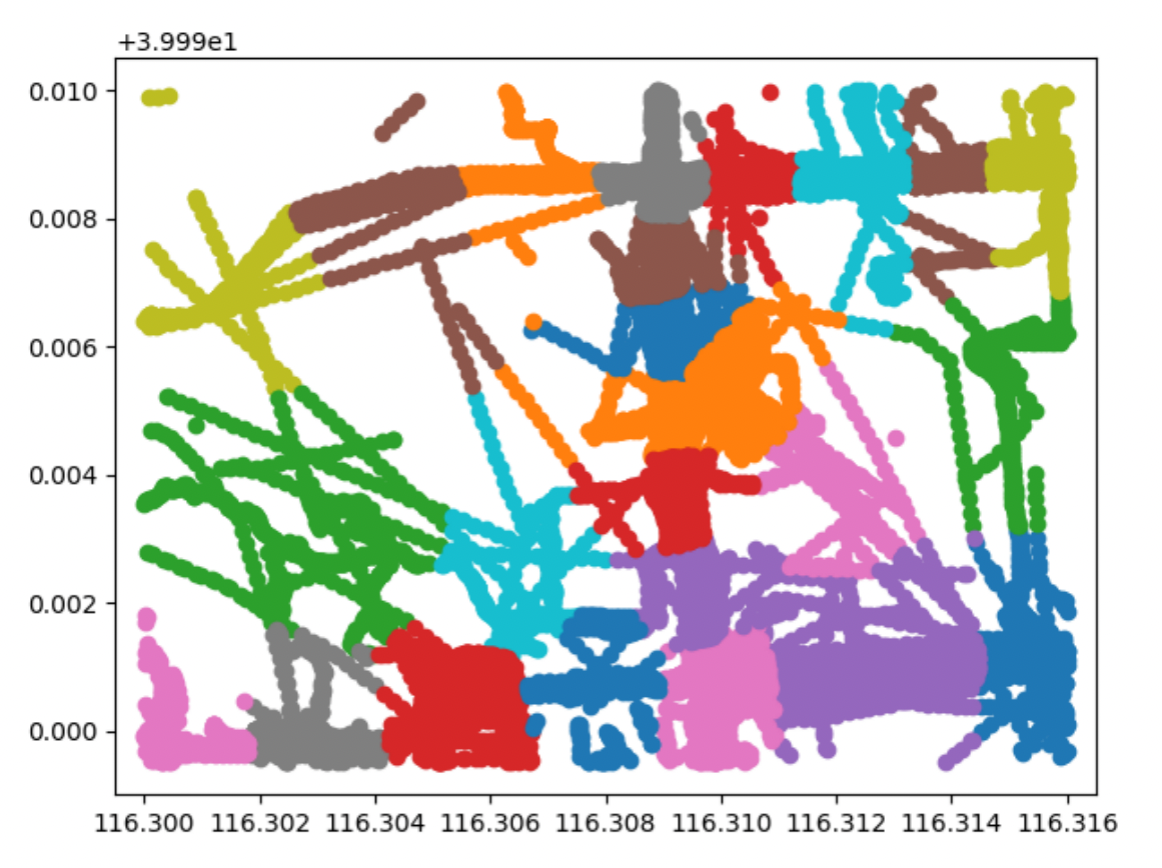}
	\caption{Trajectory point set clustering results}
	\Description{Trajectory set road network diagram, trajectory points are represented by circles, each cluster has different color.}
\end{figure}

\begin{figure}[h]
	\centering
	\includegraphics[width=\linewidth]{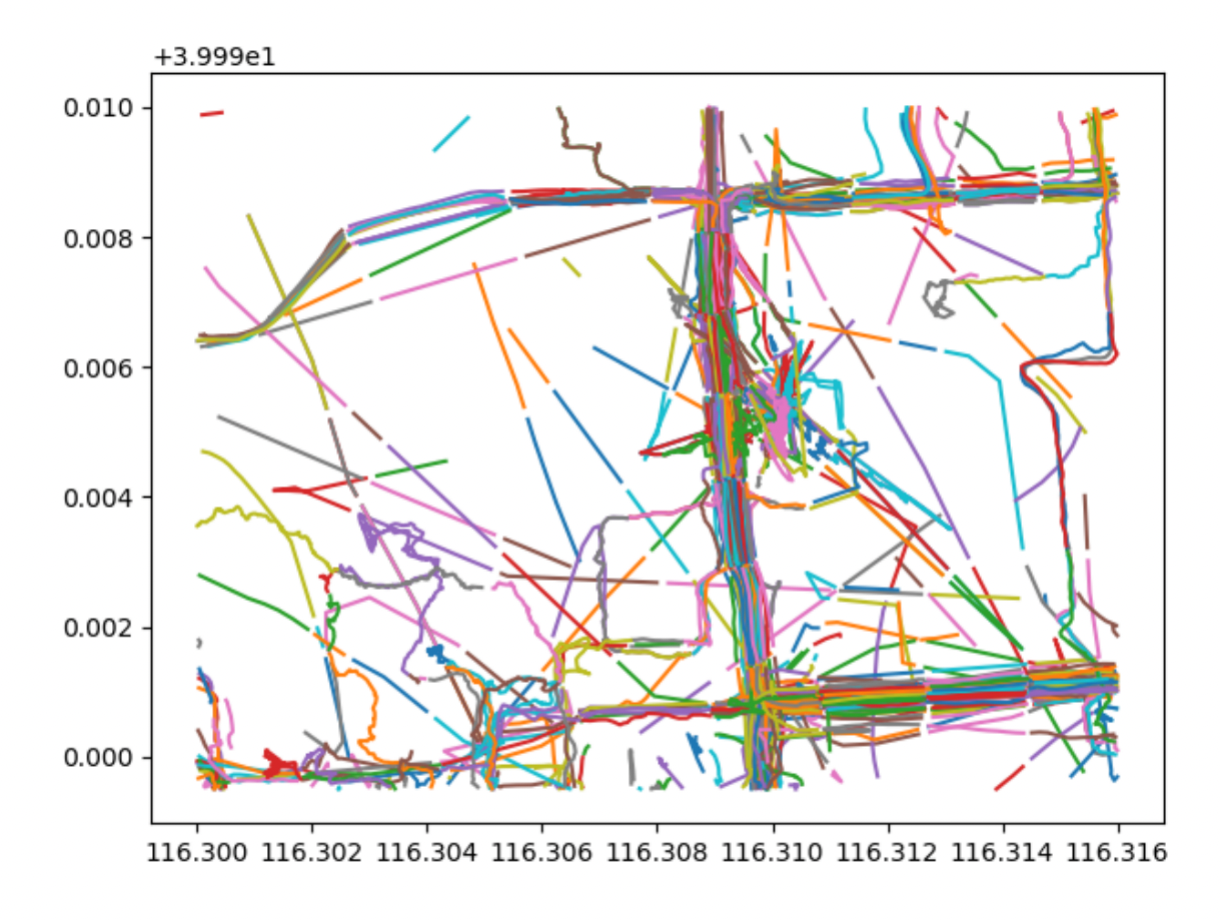}
	\caption{Road network mapping of the segmented trajectory set}
	\Description{Trajectory set road network diagram which after segmented, different trajectories are marked with different colors, and trajectory points are connected to trajectory points by lines}
\end{figure}

\begin{figure*}[ht]
	\centering
	\subfigure[Total Information Loss]{ 
		\includegraphics[width=2.2in]{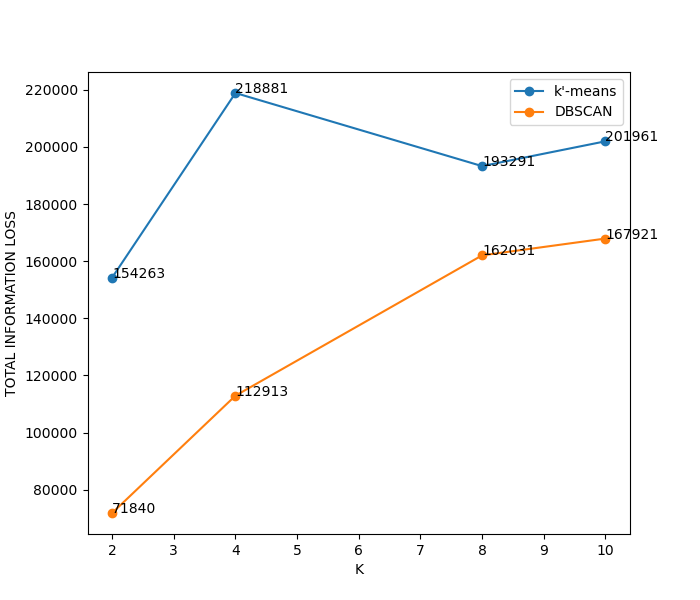} 
	} 
	\subfigure[Information Loss Per Cluster]{ 
		\includegraphics[width=2.2in]{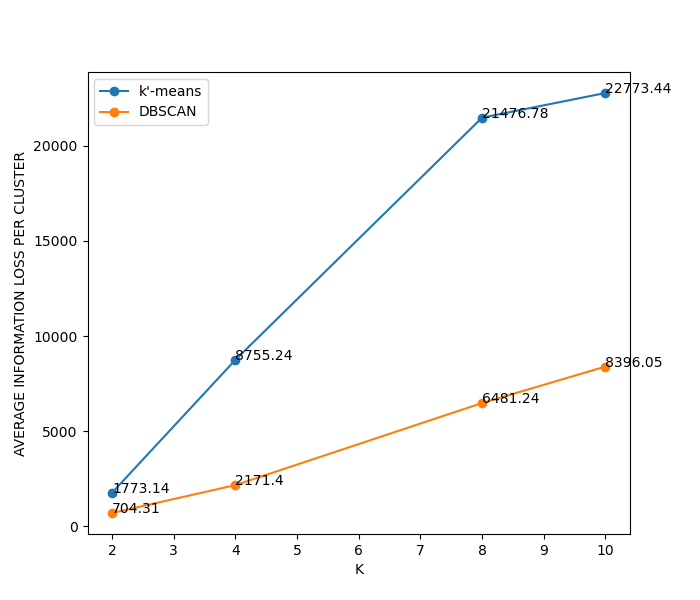} 
	}
	\subfigure[Execution Time]{ 
		\includegraphics[width=2.2in]{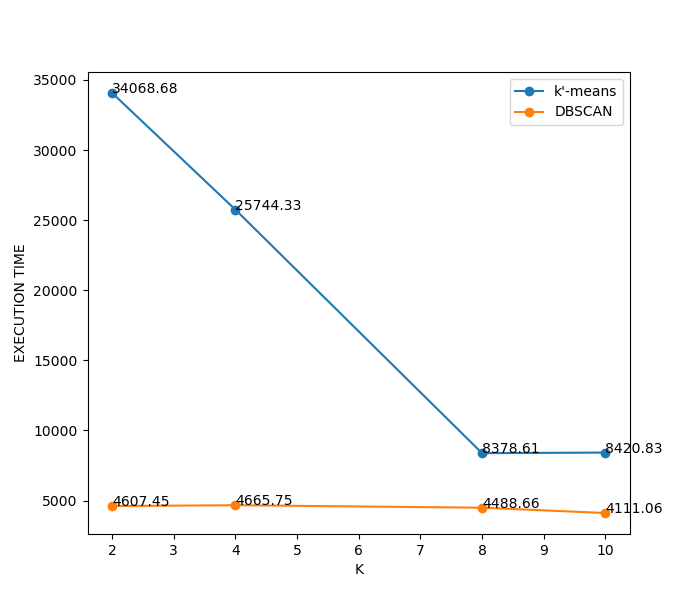} 
	} 
	\caption{Comparison of the three performances of the two trajectory clustering algorithms without segmentation preprocessing at different $k$ values}
	\Description{There are two lines of line graph, two lines represent the performance of k-means and DBSCAN in total information loss, information loss per cluster and execution time before trajectory segmentation, the orange line is DBSCAN and the blue line is k'-means.}
\end{figure*}

\begin{figure*}[h]
	\centering
	\subfigure[Total Information Loss]{ 
		\includegraphics[width=2.2in]{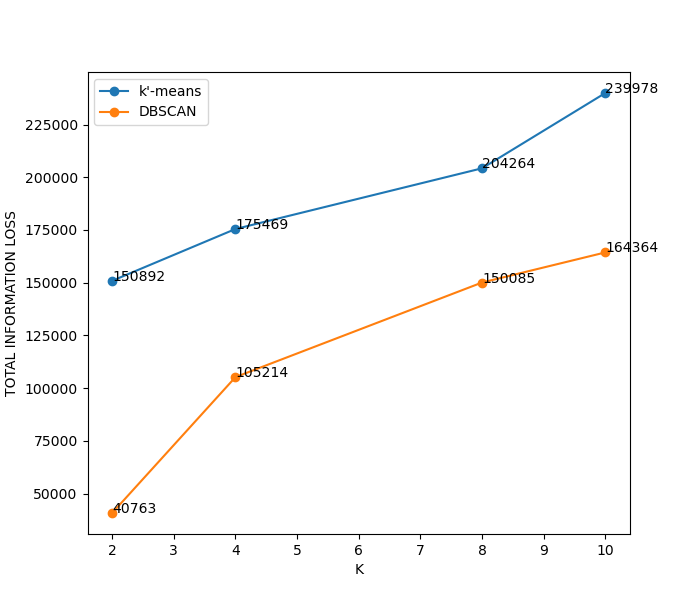} 
	} 
	\subfigure[Information Loss Per Cluster]{ 
		\includegraphics[width=2.2in]{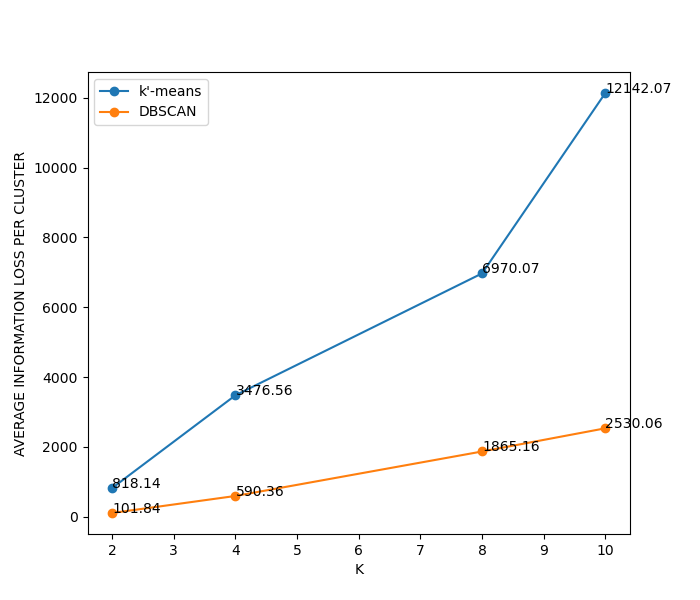} 
	}
	\subfigure[Execution Time]{ 
		\includegraphics[width=2.2in]{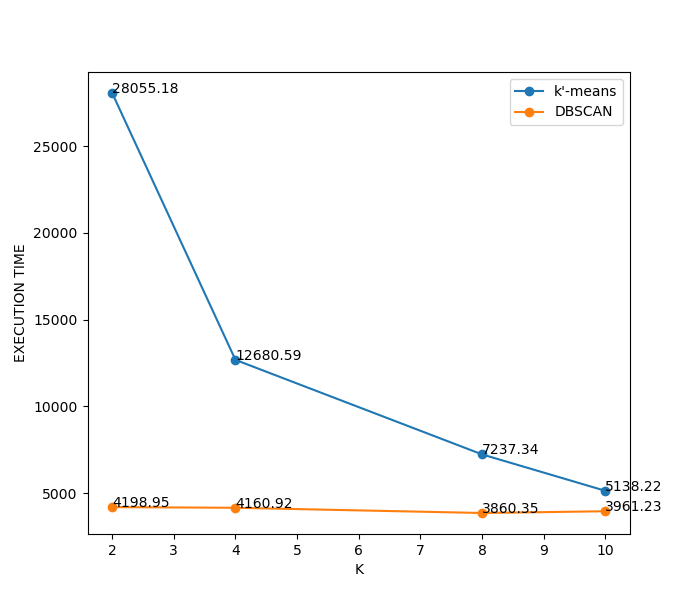} 
	} 
	\caption{After segmentation preprocessing, the three performance comparisons of the two trajectory clustering algorithms at different $k$ values}
	\Description{There are two lines of line graph, two lines represent the performance of k-means and DBSCAN in total information loss, information loss per cluster and execution time after trajectory segmentation, the orange line is DBSCAN and the blue line is k'-means.}
\end{figure*}

\subsection{Experimental Process}
In the process of realizing k-anonymity privacy protection for trajectory datasets, in order to enhance the data utility of the final anonymous trajectory set, the model in this paper preprocesses the original trajectory set by segmenting the trajectories based on point density. In the preprocessing, firstly, auxiliary points used to reflect the spatial distribution of trajectories need to be generated on the trajectories, and then all the points in the trajectory set are k-means clustered, and finally, the trajectories are partitioned according to the clustering of the trajectory point set. If the adjacent points on a trajectory are grouped into different clusters, the trajectory is segmented. Figure 5 shows the clustering results of dividing all points in the trajectory set into 27 clusters by k-means after adding auxiliary points, and different colors are used in the figure to identify the different clusters to which the point set belongs. Based on the clustering in Figure 6, the segmented trajectory set generated by segmenting the original trajectory set is shown in Figure 7, and again, the trajectories are distinguished from each other by color. Due to the randomness of the k-means algorithm in selecting the initial clustering centers, the segmentation results usually vary from experiment to experiment, and the number of segmented trajectories is in the range of about 1200 to 1450, as obtained from a large number of reliable repeated experiments. In the experiments in Figures 6 and 7, the number of segmented trajectories increases from 270 to 1372.

In order to make the final published trajectory dataset resistant to re-identification attacks, the trajectory privacy-preserving model needs to anonymize the trajectory set according to the selected k-values in k-anonymity criterion. In this process, the DGH tree generalization model of the trajectory set needs to be established first, i.e., the corresponding coordinate values of the trajectory points are represented by the numbers of the leaf nodes on the latitude and longitude DGH trees. Then the iterative k'-means algorithm and the adaptive DBSCAN algorithm cluster the trajectories, respectively. Finally, the clusters of trajectories formed by the clusters are generalized to obtain the trajectory dataset conforming to k-anonymity and the corresponding loss of generalization information.

\begin{table*}[ht]
	\caption{KPDP performance with Partition model and without Partition model}
	\begin{tabular}{|ccccc|ccccc|}
		\hline
		\multicolumn{5}{|c|}{Total Information Loss} & \multicolumn{5}{c|}{Average Information Loss Per Cluster} \\ \hline
		\multicolumn{1}{|c|}{\begin{tabular}[c]{@{}c@{}}k \\ value\end{tabular}} & \multicolumn{1}{c|}{\begin{tabular}[c]{@{}c@{}}clustering \\ algorithm\end{tabular}} & \multicolumn{1}{c|}{\begin{tabular}[c]{@{}c@{}}without \\ Parition \\ model\end{tabular}} & \multicolumn{1}{c|}{\begin{tabular}[c]{@{}c@{}}with\\ partition \\ model\end{tabular}} & reduction(\%) & \multicolumn{1}{c|}{\begin{tabular}[c]{@{}c@{}}k \\ value\end{tabular}} & \multicolumn{1}{c|}{\begin{tabular}[c]{@{}c@{}}clustering \\ algorithm\end{tabular}} & \multicolumn{1}{c|}{\begin{tabular}[c]{@{}c@{}}without \\ Parition \\ model\end{tabular}} & \multicolumn{1}{c|}{\begin{tabular}[c]{@{}c@{}}with\\ partition \\ model\end{tabular}} & reduction(\%) \\ \hline
		\multicolumn{1}{|c|}{\multirow{2}{*}{k=2}} & \multicolumn{1}{c|}{k‘-means} & \multicolumn{1}{c|}{154263} & \multicolumn{1}{c|}{150892} & 2.19 & \multicolumn{1}{c|}{\multirow{2}{*}{k=2}} & \multicolumn{1}{c|}{k’-means} & \multicolumn{1}{c|}{1773.14} & \multicolumn{1}{c|}{818.14} & 53.86 \\ \cline{2-5} \cline{7-10} 
		\multicolumn{1}{|c|}{} & \multicolumn{1}{c|}{DBSCAN} & \multicolumn{1}{c|}{71840} & \multicolumn{1}{c|}{40763} & 43.26 & \multicolumn{1}{c|}{} & \multicolumn{1}{c|}{DBSCAN} & \multicolumn{1}{c|}{704.31} & \multicolumn{1}{c|}{101.84} & 85.54 \\ \hline
		\multicolumn{1}{|c|}{\multirow{2}{*}{k=4}} & \multicolumn{1}{c|}{k‘-means} & \multicolumn{1}{c|}{218881} & \multicolumn{1}{c|}{175469} & 19.83 & \multicolumn{1}{c|}{\multirow{2}{*}{k=4}} & \multicolumn{1}{c|}{k’-means} & \multicolumn{1}{c|}{8755.24} & \multicolumn{1}{c|}{3476.56} & 60.29 \\ \cline{2-5} \cline{7-10} 
		\multicolumn{1}{|c|}{} & \multicolumn{1}{c|}{DBSCAN} & \multicolumn{1}{c|}{112913} & \multicolumn{1}{c|}{105214} & 6.82 & \multicolumn{1}{c|}{} & \multicolumn{1}{c|}{DBSCAN} & \multicolumn{1}{c|}{2171.4} & \multicolumn{1}{c|}{590.36} & 72.81 \\ \hline
		\multicolumn{1}{|c|}{\multirow{2}{*}{k=8}} & \multicolumn{1}{c|}{k‘-means} & \multicolumn{1}{c|}{193291} & \multicolumn{1}{c|}{204264} & -5.68 & \multicolumn{1}{c|}{\multirow{2}{*}{k=8}} & \multicolumn{1}{c|}{k’-means} & \multicolumn{1}{c|}{21476.78} & \multicolumn{1}{c|}{6970.07} & 67.55 \\ \cline{2-5} \cline{7-10} 
		\multicolumn{1}{|c|}{} & \multicolumn{1}{c|}{DBSCAN} & \multicolumn{1}{c|}{162031} & \multicolumn{1}{c|}{150085} & 7.37 & \multicolumn{1}{c|}{} & \multicolumn{1}{c|}{DBSCAN} & \multicolumn{1}{c|}{6481.24} & \multicolumn{1}{c|}{1865.16} & 71.22 \\ \hline
		\multicolumn{1}{|c|}{\multirow{2}{*}{k=10}} & \multicolumn{1}{c|}{k‘-means} & \multicolumn{1}{c|}{201961} & \multicolumn{1}{c|}{239978} & -18.82 & \multicolumn{1}{c|}{\multirow{2}{*}{k=10}} & \multicolumn{1}{c|}{k’-means} & \multicolumn{1}{c|}{22773.44} & \multicolumn{1}{c|}{12142.007} & 46.68 \\ \cline{2-5} \cline{7-10} 
		\multicolumn{1}{|c|}{} & \multicolumn{1}{c|}{DBSCAN} & \multicolumn{1}{c|}{167921} & \multicolumn{1}{c|}{164364} & 2.12 & \multicolumn{1}{c|}{} & \multicolumn{1}{c|}{DBSCAN} & \multicolumn{1}{c|}{8396.05} & \multicolumn{1}{c|}{2530.06} & 69.87 \\ \hline
	\end{tabular}
\end{table*}

\subsection{Analysis of Experimental Results}
In the experiments on trajectory privacy preservation against re-identification attacks, this paper will compare and evaluate two trajectory clustering algorithms with and without segment preprocessing in three aspects: total information loss, average information loss per cluster, and execution time.

Figure 8 shows the comparison of the values of the three metrics obtained by the iterative k'-means algorithm and the adaptive DBSCAN algorithm for different k-anonymity metrics without trajectory segmentation preprocessing during the experiment, where the k-anonymity criterion takes the $k$ of 2, 4, 8 and 10. 

As shown in Figure 8(a), the total generalized information loss of the trajectory set increases with the increases of $k$. In contrast, the protection model for trajectory, which is clustering by the adaptive DBSCAN algorithm, produces lower information loss than the iterative k'-means algorithm at all four values. 

As shown in Figure 8(b), the average generalized information loss per cluster of the trajectory set also increases with increasing $k$. The information loss generated by the iterative k'-means algorithm is two to four times higher than that of the adaptive DBSCAN algorithm. The difference becomes more pronounced as the values increase. 

The execution time of the two trajectory clustering algorithms in the model is shown in Figure 8(c), with a decreasing trend of the algorithm execution time when increasing. In the experiments for each value, the execution time of the adaptive DBSCAN algorithm is stable within 5000 seconds, while the execution time of the iterative k'-means algorithm is much higher than the other algorithm for values 2 and 4, and relatively lower and smoother for values 8 and 10, but still higher than the other algorithm.

In Figure 9, a comparison of the values of the three metrics obtained by the iterative k'-means algorithm and the adaptive DBSCAN algorithm with different k-anonymity criteria for the dataset preprocessed by trajectory segmentation at the time of the experiment is shown, where the k-anonymity criteria take the values of 2, 4, 8 and 10. 

Similar to the overall trend and comparison in Figure 8, the adaptive DBSCAN algorithm outperforms the iterative k'-means algorithm in three aspects: total generalized information loss (Figure 9(a)), average information loss per cluster (Figure 9(b)), and execution time (Figure 9(c)). Overall, the total generalized information loss and the average information loss per cluster of both trajectory clustering algorithms subsequently increase with the increase of $k$, and the execution time decreases as the value of $k$ increases gradually.

In contrast, compared with Fig 8, in both the adaptive DBSCAN algorithm and the iterative k'-means algorithm, the total information loss and average information loss per cluster of the final generated anonymized dataset after segmentation preprocessing by the partition model are relatively small, and the consumed execution time is also reduced to different degrees. Especially it is evident in the per-cluster average information loss metric of the adaptive DBSCAN algorithm, which can be obtained from Figure 9(b), that the information loss per cluster obtained by the adaptive DBSCAN algorithm decreases by about 86\%, 73\%, 71\% and 70\%, respectively, when the values of 2, 4, 8 and 10 are taken. Compared with the trajectory set without segmentation in Figure 8(b). 

Such a decrease is because the preprocessing of the partition model can make the clustered trajectories closer within the clusters. Besides, the smaller the value $k$, the larger the number of clusters after segmentation, and the closer the trajectories that make up the clusters will be. The more obvious is the effect of the partition model in reducing the information loss of generalization within the clusters.

The experimental results shown in Figure 8 and Figure 9 are consistent with the expectations of KPDP design in this paper. For the case where the generalized information loss increases with the value of $k$, this is because an increase in the value of $k$ directly leads to an increase in the number of trajectories in each cluster, resulting in a more extensive total information loss and average information loss per cluster. The superior performance of the adaptive DBSCAN algorithm in the three metrics is attributed to the ability of the algorithm to cluster trajectories close to each other more scientifically and efficiently than the iterative k'-means algorithm, with lower time complexity. 

According to a series of experiments, it is proved that the way of adjusting the cluster centers in the k'-means algorithm is not fully applicable to the trajectory clustering process, while the adaptive DBSCAN algorithm forms each cluster by the expansion of the density connection between trajectories, which not only reduces the information loss but also can effectively speed up the processing of the model. The effectiveness of adding a segment preprocessing step in both trajectory clustering algorithms is due to the fact that after preprocessing, the relatively long trajectories in the dataset are avoided to be aligned and combined with shorter trajectories in the trajectory clustering and generalization process, so the information loss from the final generalization is reduced. In addition, because the long trajectories in the dataset are split into relatively short trajectories, the situation that two long trajectories are aligned with each other will be significantly avoided in the alignment, so the execution time of the trajectory clustering algorithm is also shortened.

In summary, the adaptive DBSCAN algorithm and the trajectory set segmentation preprocessing step proposed in this paper to have superior performance in controlled experiments under different scenarios, validating the theoretical expectation of reducing generalization information loss and speeding up model processing when designing the model. In the privacy-preserving phase of trajectory resistance to re-identification attacks, the trajectory preprocessing and adaptive DBSCAN algorithm for trajectory clustering to form anonymous trajectory datasets in Figure 9 has significant advantages in terms of data utility and running time for each value.

\section{CONCLUSION}
In this paper, we proposed a trajectory privacy protection framework against re-identification attacks, which can effectively anonymize the spatiotemporal trajectory dataset. We innovated a point density-based trajectory segmentation preprocessing mechanism to enable accurate clustering and generalization of trajectories. Furthermore, we applied  DBSCAN in machine learning to trajectory clustering and presented the adaptive DBSCAN algorithm, which minimizes the generalization information loss to acquire higher data utility while ensuring the k-anonymity of the generated trajectory dataset. Extensive experiments on a realistic dataset also showed that there is the superiority of the short execution time of our approach compared with previous works.

\bibliographystyle{ACM-Reference-Format}
\bibliography{KPDPreference}


\begin{thebibliography}{47}


\ifx \showCODEN    \undefined \def \showCODEN     #1{\unskip}     \fi
\ifx \showDOI      \undefined \def \showDOI       #1{#1}\fi
\ifx \showISBNx    \undefined \def \showISBNx     #1{\unskip}     \fi
\ifx \showISBNxiii \undefined \def \showISBNxiii  #1{\unskip}     \fi
\ifx \showISSN     \undefined \def \showISSN      #1{\unskip}     \fi
\ifx \showLCCN     \undefined \def \showLCCN      #1{\unskip}     \fi
\ifx \shownote     \undefined \def \shownote      #1{#1}          \fi
\ifx \showarticletitle \undefined \def \showarticletitle #1{#1}   \fi
\ifx \showURL      \undefined \def \showURL       {\relax}        \fi
\providecommand\bibfield[2]{#2}
\providecommand\bibinfo[2]{#2}
\providecommand\natexlab[1]{#1}
\providecommand\showeprint[2][]{arXiv:#2}

\bibitem[Anjum and Raschia(2017)]%
        {anjum2017banga}
\bibfield{author}{\bibinfo{person}{Adeel Anjum} {and}
  \bibinfo{person}{Guillaume Raschia}.} \bibinfo{year}{2017}\natexlab{}.
\newblock \showarticletitle{BangA: an efficient and flexible
  generalization-based algorithm for privacy preserving data publication}.
\newblock \bibinfo{journal}{\emph{Computers}} \bibinfo{volume}{6},
  \bibinfo{number}{1} (\bibinfo{year}{2017}), \bibinfo{pages}{1}.
\newblock


\bibitem[Beresford and Stajano(2004)]%
        {beresford2004mix}
\bibfield{author}{\bibinfo{person}{Alastair~R Beresford} {and}
  \bibinfo{person}{Frank Stajano}.} \bibinfo{year}{2004}\natexlab{}.
\newblock \showarticletitle{Mix zones: User privacy in location-aware
  services}. In \bibinfo{booktitle}{\emph{IEEE Annual conference on pervasive
  computing and communications workshops, 2004. Proceedings of the Second}}.
  IEEE, \bibinfo{pages}{127--131}.
\newblock


\bibitem[Bettini et~al\mbox{.}(2009)]%
        {bettini2009anonymity}
\bibfield{author}{\bibinfo{person}{Claudio Bettini}, \bibinfo{person}{Sergio
  Mascetti}, \bibinfo{person}{X~Sean Wang}, \bibinfo{person}{Dario Freni},
  {and} \bibinfo{person}{Sushil Jajodia}.} \bibinfo{year}{2009}\natexlab{}.
\newblock \showarticletitle{Anonymity and historical-anonymity in
  location-based services}.
\newblock \bibinfo{journal}{\emph{Privacy in location-based applications:
  research issues and emerging trends}} (\bibinfo{year}{2009}),
  \bibinfo{pages}{1--30}.
\newblock


\bibitem[Campan et~al\mbox{.}(2011)]%
        {campan2011fly}
\bibfield{author}{\bibinfo{person}{Alina Campan}, \bibinfo{person}{Nicholas
  Cooper}, {and} \bibinfo{person}{Traian~Marius Truta}.}
  \bibinfo{year}{2011}\natexlab{}.
\newblock \showarticletitle{On-the-fly generalization hierarchies for numerical
  attributes revisited}. In \bibinfo{booktitle}{\emph{Secure Data Management:
  8th VLDB Workshop, SDM 2011, Seattle, WA, USA, September 2, 2011, Proceedings
  8}}. Springer, \bibinfo{pages}{18--32}.
\newblock


\bibitem[Chen et~al\mbox{.}(2017)]%
        {chen2017cmsa}
\bibfield{author}{\bibinfo{person}{Xi Chen}, \bibinfo{person}{Chen Wang},
  \bibinfo{person}{Shanjiang Tang}, \bibinfo{person}{Ce Yu}, {and}
  \bibinfo{person}{Quan Zou}.} \bibinfo{year}{2017}\natexlab{}.
\newblock \showarticletitle{CMSA: a heterogeneous CPU/GPU computing system for
  multiple similar RNA/DNA sequence alignment}.
\newblock \bibinfo{journal}{\emph{BMC bioinformatics}}  \bibinfo{volume}{18}
  (\bibinfo{year}{2017}), \bibinfo{pages}{1--10}.
\newblock


\bibitem[Chow and Golle(2009)]%
        {chow2009faking}
\bibfield{author}{\bibinfo{person}{Richard Chow} {and}
  \bibinfo{person}{Philippe Golle}.} \bibinfo{year}{2009}\natexlab{}.
\newblock \showarticletitle{Faking contextual data for fun, profit, and
  privacy}. In \bibinfo{booktitle}{\emph{Proceedings of the 8th ACM workshop on
  Privacy in the electronic society}}. \bibinfo{pages}{105--108}.
\newblock


\bibitem[Chowdhury and Garai(2017)]%
        {chowdhury2017review}
\bibfield{author}{\bibinfo{person}{Biswanath Chowdhury} {and}
  \bibinfo{person}{Gautam Garai}.} \bibinfo{year}{2017}\natexlab{}.
\newblock \showarticletitle{A review on multiple sequence alignment from the
  perspective of genetic algorithm}.
\newblock \bibinfo{journal}{\emph{Genomics}} \bibinfo{volume}{109},
  \bibinfo{number}{5-6} (\bibinfo{year}{2017}), \bibinfo{pages}{419--431}.
\newblock


\bibitem[Cicek et~al\mbox{.}(2014)]%
        {cicek2014ensuring}
\bibfield{author}{\bibinfo{person}{A~Ercument Cicek},
  \bibinfo{person}{Mehmet~Ercan Nergiz}, {and} \bibinfo{person}{Yucel Saygin}.}
  \bibinfo{year}{2014}\natexlab{}.
\newblock \showarticletitle{Ensuring location diversity in privacy-preserving
  spatio-temporal data publishing}.
\newblock \bibinfo{journal}{\emph{The VLDB Journal}} \bibinfo{volume}{23},
  \bibinfo{number}{4} (\bibinfo{year}{2014}), \bibinfo{pages}{609--625}.
\newblock


\bibitem[Ding(2015)]%
        {ding2015trajectory}
\bibfield{author}{\bibinfo{person}{Jiaxin Ding}.}
  \bibinfo{year}{2015}\natexlab{}.
\newblock \showarticletitle{Trajectory mining, representation and privacy
  protection}. In \bibinfo{booktitle}{\emph{Proceedings of the 2nd ACM
  SIGSPATIAL PhD Workshop}}. \bibinfo{pages}{1--4}.
\newblock


\bibitem[Do et~al\mbox{.}(2016)]%
        {do2016another}
\bibfield{author}{\bibinfo{person}{Hyo~Jin Do}, \bibinfo{person}{Young-Seob
  Jeong}, \bibinfo{person}{Ho-Jin Choi}, {and} \bibinfo{person}{Kwangjo Kim}.}
  \bibinfo{year}{2016}\natexlab{}.
\newblock \showarticletitle{Another dummy generation technique in
  location-based services}. In \bibinfo{booktitle}{\emph{2016 International
  Conference on Big Data and Smart Computing (BigComp)}}. IEEE,
  \bibinfo{pages}{532--538}.
\newblock


\bibitem[Ester et~al\mbox{.}(1996)]%
        {ester1996density}
\bibfield{author}{\bibinfo{person}{Martin Ester}, \bibinfo{person}{Hans-Peter
  Kriegel}, \bibinfo{person}{J{\"o}rg Sander}, \bibinfo{person}{Xiaowei Xu},
  {et~al\mbox{.}}} \bibinfo{year}{1996}\natexlab{}.
\newblock \showarticletitle{A density-based algorithm for discovering clusters
  in large spatial databases with noise.}. In \bibinfo{booktitle}{\emph{kdd}},
  Vol.~\bibinfo{volume}{96}. \bibinfo{pages}{226--231}.
\newblock


\bibitem[Fung et~al\mbox{.}(2005)]%
        {fung2005top}
\bibfield{author}{\bibinfo{person}{Benjamin~CM Fung}, \bibinfo{person}{Ke
  Wang}, {and} \bibinfo{person}{Philip~S Yu}.} \bibinfo{year}{2005}\natexlab{}.
\newblock \showarticletitle{Top-down specialization for information and privacy
  preservation}. In \bibinfo{booktitle}{\emph{21st international conference on
  data engineering (ICDE'05)}}. IEEE, \bibinfo{pages}{205--216}.
\newblock


\bibitem[Ghinita et~al\mbox{.}(2009)]%
        {ghinita2009preventing}
\bibfield{author}{\bibinfo{person}{Gabriel Ghinita},
  \bibinfo{person}{Maria~Luisa Damiani}, \bibinfo{person}{Claudio Silvestri},
  {and} \bibinfo{person}{Elisa Bertino}.} \bibinfo{year}{2009}\natexlab{}.
\newblock \showarticletitle{Preventing velocity-based linkage attacks in
  location-aware applications}. In \bibinfo{booktitle}{\emph{Proceedings of the
  17th ACM SIGSPATIAL international conference on advances in geographic
  information systems}}. \bibinfo{pages}{246--255}.
\newblock


\bibitem[Gramaglia et~al\mbox{.}(2017)]%
        {DBLP:journals/corr/GramagliaFTB17}
\bibfield{author}{\bibinfo{person}{Marco Gramaglia}, \bibinfo{person}{Marco
  Fiore}, \bibinfo{person}{Alberto Tarable}, {and} \bibinfo{person}{Albert
  Banchs}.} \bibinfo{year}{2017}\natexlab{}.
\newblock \showarticletitle{k\({}^{\mbox{{\(\tau\)},
  {\(\epsilon\)}}}\)-anonymity: Towards Privacy-Preserving Publishing of
  Spatiotemporal Trajectory Data}.
\newblock \bibinfo{journal}{\emph{CoRR}}  \bibinfo{volume}{abs/1701.02243}
  (\bibinfo{year}{2017}).
\newblock
\showeprint[arXiv]{1701.02243}
\urldef\tempurl%
\url{http://arxiv.org/abs/1701.02243}
\showURL{%
\tempurl}


\bibitem[Gruteser and Grunwald(2003)]%
        {gruteser2003anonymous}
\bibfield{author}{\bibinfo{person}{Marco Gruteser} {and} \bibinfo{person}{Dirk
  Grunwald}.} \bibinfo{year}{2003}\natexlab{}.
\newblock \showarticletitle{Anonymous usage of location-based services through
  spatial and temporal cloaking}. In \bibinfo{booktitle}{\emph{Proceedings of
  the 1st international conference on Mobile systems, applications and
  services}}. \bibinfo{pages}{31--42}.
\newblock


\bibitem[Iyengar(2002)]%
        {iyengar2002transforming}
\bibfield{author}{\bibinfo{person}{Vijay~S Iyengar}.}
  \bibinfo{year}{2002}\natexlab{}.
\newblock \showarticletitle{Transforming data to satisfy privacy constraints}.
  In \bibinfo{booktitle}{\emph{Proceedings of the eighth ACM SIGKDD
  international conference on Knowledge discovery and data mining}}.
  \bibinfo{pages}{279--288}.
\newblock


\bibitem[Jiang et~al\mbox{.}(2013)]%
        {jiang2013publishing}
\bibfield{author}{\bibinfo{person}{Kaifeng Jiang}, \bibinfo{person}{Dongxu
  Shao}, \bibinfo{person}{St{\'e}phane Bressan}, \bibinfo{person}{Thomas
  Kister}, {and} \bibinfo{person}{Kian-Lee Tan}.}
  \bibinfo{year}{2013}\natexlab{}.
\newblock \showarticletitle{Publishing trajectories with differential privacy
  guarantees}. In \bibinfo{booktitle}{\emph{Proceedings of the 25th
  International conference on scientific and statistical database management}}.
  \bibinfo{pages}{1--12}.
\newblock


\bibitem[Krumm(2007)]%
        {krumm2007inference}
\bibfield{author}{\bibinfo{person}{John Krumm}.}
  \bibinfo{year}{2007}\natexlab{}.
\newblock \showarticletitle{Inference attacks on location tracks}. In
  \bibinfo{booktitle}{\emph{Pervasive Computing: 5th International Conference,
  PERVASIVE 2007, Toronto, Canada, May 13-16, 2007. Proceedings 5}}. Springer,
  \bibinfo{pages}{127--143}.
\newblock


\bibitem[Le et~al\mbox{.}(2017)]%
        {le2017protein}
\bibfield{author}{\bibinfo{person}{Quan Le}, \bibinfo{person}{Fabian Sievers},
  {and} \bibinfo{person}{Desmond~G Higgins}.} \bibinfo{year}{2017}\natexlab{}.
\newblock \showarticletitle{Protein multiple sequence alignment benchmarking
  through secondary structure prediction}.
\newblock \bibinfo{journal}{\emph{Bioinformatics}} \bibinfo{volume}{33},
  \bibinfo{number}{9} (\bibinfo{year}{2017}), \bibinfo{pages}{1331--1337}.
\newblock


\bibitem[Lee et~al\mbox{.}(2007)]%
        {lee2007trajectory}
\bibfield{author}{\bibinfo{person}{Jae-Gil Lee}, \bibinfo{person}{Jiawei Han},
  {and} \bibinfo{person}{Kyu-Young Whang}.} \bibinfo{year}{2007}\natexlab{}.
\newblock \showarticletitle{Trajectory clustering: a partition-and-group
  framework}. In \bibinfo{booktitle}{\emph{Proceedings of the 2007 ACM SIGMOD
  international conference on Management of data}}. \bibinfo{pages}{593--604}.
\newblock


\bibitem[LeFevre et~al\mbox{.}(2005)]%
        {lefevre2005incognito}
\bibfield{author}{\bibinfo{person}{Kristen LeFevre}, \bibinfo{person}{David~J
  DeWitt}, {and} \bibinfo{person}{Raghu Ramakrishnan}.}
  \bibinfo{year}{2005}\natexlab{}.
\newblock \showarticletitle{Incognito: Efficient full-domain k-anonymity}. In
  \bibinfo{booktitle}{\emph{Proceedings of the 2005 ACM SIGMOD international
  conference on Management of data}}. \bibinfo{pages}{49--60}.
\newblock


\bibitem[LeFevre et~al\mbox{.}(2006)]%
        {lefevre2006mondrian}
\bibfield{author}{\bibinfo{person}{Kristen LeFevre}, \bibinfo{person}{David~J
  DeWitt}, {and} \bibinfo{person}{Raghu Ramakrishnan}.}
  \bibinfo{year}{2006}\natexlab{}.
\newblock \showarticletitle{Mondrian multidimensional k-anonymity}. In
  \bibinfo{booktitle}{\emph{22nd International conference on data engineering
  (ICDE'06)}}. IEEE, \bibinfo{pages}{25--25}.
\newblock


\bibitem[Li et~al\mbox{.}(2016)]%
        {li2016privacy}
\bibfield{author}{\bibinfo{person}{Huaxin Li}, \bibinfo{person}{Haojin Zhu},
  \bibinfo{person}{Suguo Du}, \bibinfo{person}{Xiaohui Liang}, {and}
  \bibinfo{person}{Xuemin Shen}.} \bibinfo{year}{2016}\natexlab{}.
\newblock \showarticletitle{Privacy leakage of location sharing in mobile
  social networks: Attacks and defense}.
\newblock \bibinfo{journal}{\emph{IEEE Transactions on Dependable and Secure
  Computing}} \bibinfo{volume}{15}, \bibinfo{number}{4} (\bibinfo{year}{2016}),
  \bibinfo{pages}{646--660}.
\newblock


\bibitem[Liu et~al\mbox{.}(2018)]%
        {liu2018location}
\bibfield{author}{\bibinfo{person}{Bo Liu}, \bibinfo{person}{Wanlei Zhou},
  \bibinfo{person}{Tianqing Zhu}, \bibinfo{person}{Longxiang Gao}, {and}
  \bibinfo{person}{Yong Xiang}.} \bibinfo{year}{2018}\natexlab{}.
\newblock \showarticletitle{Location privacy and its applications: A systematic
  study}.
\newblock \bibinfo{journal}{\emph{IEEE access}}  \bibinfo{volume}{6}
  (\bibinfo{year}{2018}), \bibinfo{pages}{17606--17624}.
\newblock


\bibitem[Machanavajjhala et~al\mbox{.}(2007)]%
        {machanavajjhala2007diversity}
\bibfield{author}{\bibinfo{person}{Ashwin Machanavajjhala},
  \bibinfo{person}{Daniel Kifer}, \bibinfo{person}{Johannes Gehrke}, {and}
  \bibinfo{person}{Muthuramakrishnan Venkitasubramaniam}.}
  \bibinfo{year}{2007}\natexlab{}.
\newblock \showarticletitle{l-diversity: Privacy beyond k-anonymity}.
\newblock \bibinfo{journal}{\emph{ACM Transactions on Knowledge Discovery from
  Data (TKDD)}} \bibinfo{volume}{1}, \bibinfo{number}{1}
  (\bibinfo{year}{2007}), \bibinfo{pages}{3--es}.
\newblock


\bibitem[MacQueen(1967)]%
        {macqueen1967classification}
\bibfield{author}{\bibinfo{person}{J MacQueen}.}
  \bibinfo{year}{1967}\natexlab{}.
\newblock \showarticletitle{Classification and analysis of multivariate
  observations}. In \bibinfo{booktitle}{\emph{5th Berkeley Symp. Math. Statist.
  Probability}}. University of California Los Angeles LA USA,
  \bibinfo{pages}{281--297}.
\newblock


\bibitem[Mokbel(2007)]%
        {mokbel2007privacy}
\bibfield{author}{\bibinfo{person}{Mohamed~F Mokbel}.}
  \bibinfo{year}{2007}\natexlab{}.
\newblock \showarticletitle{Privacy in location-based services:
  State-of-the-art and research directions}. In \bibinfo{booktitle}{\emph{2007
  International Conference on Mobile Data Management}}. IEEE Computer Society,
  \bibinfo{pages}{228--228}.
\newblock


\bibitem[Naghizade et~al\mbox{.}(2014)]%
        {naghizade2014protection}
\bibfield{author}{\bibinfo{person}{Elham Naghizade}, \bibinfo{person}{Lars
  Kulik}, {and} \bibinfo{person}{Egemen Tanin}.}
  \bibinfo{year}{2014}\natexlab{}.
\newblock \showarticletitle{Protection of sensitive trajectory datasets through
  spatial and temporal exchange}. In \bibinfo{booktitle}{\emph{Proceedings of
  the 26th International Conference on Scientific and Statistical Database
  Management}}. \bibinfo{pages}{1--4}.
\newblock


\bibitem[Samarati(2001)]%
        {samarati2001protecting}
\bibfield{author}{\bibinfo{person}{Pierangela Samarati}.}
  \bibinfo{year}{2001}\natexlab{}.
\newblock \showarticletitle{Protecting respondents identities in microdata
  release}.
\newblock \bibinfo{journal}{\emph{IEEE transactions on Knowledge and Data
  Engineering}} \bibinfo{volume}{13}, \bibinfo{number}{6}
  (\bibinfo{year}{2001}), \bibinfo{pages}{1010--1027}.
\newblock


\bibitem[Samarati and Sweeney(1998)]%
        {samarati1998protecting}
\bibfield{author}{\bibinfo{person}{Pierangela Samarati} {and}
  \bibinfo{person}{Latanya Sweeney}.} \bibinfo{year}{1998}\natexlab{}.
\newblock \showarticletitle{Protecting privacy when disclosing information:
  k-anonymity and its enforcement through generalization and suppression}.
\newblock  (\bibinfo{year}{1998}).
\newblock


\bibitem[Shaham et~al\mbox{.}(2020)]%
        {shaham2020privacy}
\bibfield{author}{\bibinfo{person}{Sina Shaham}, \bibinfo{person}{Ming Ding},
  \bibinfo{person}{Bo Liu}, \bibinfo{person}{Shuping Dang},
  \bibinfo{person}{Zihuai Lin}, {and} \bibinfo{person}{Jun Li}.}
  \bibinfo{year}{2020}\natexlab{}.
\newblock \showarticletitle{Privacy preserving location data publishing: A
  machine learning approach}.
\newblock \bibinfo{journal}{\emph{IEEE Transactions on Knowledge and Data
  Engineering}} \bibinfo{volume}{33}, \bibinfo{number}{9}
  (\bibinfo{year}{2020}), \bibinfo{pages}{3270--3283}.
\newblock


\bibitem[Shokri et~al\mbox{.}(2010)]%
        {shokri2010unified}
\bibfield{author}{\bibinfo{person}{Reza Shokri}, \bibinfo{person}{Julien
  Freudiger}, {and} \bibinfo{person}{Jean-Pierre Hubaux}.}
  \bibinfo{year}{2010}\natexlab{}.
\newblock \bibinfo{booktitle}{\emph{A unified framework for location privacy}}.
\newblock \bibinfo{type}{{T}echnical {R}eport}.
\newblock


\bibitem[Shokri et~al\mbox{.}(2011)]%
        {shokri2011quantifying}
\bibfield{author}{\bibinfo{person}{Reza Shokri}, \bibinfo{person}{George
  Theodorakopoulos}, \bibinfo{person}{Jean-Yves Le~Boudec}, {and}
  \bibinfo{person}{Jean-Pierre Hubaux}.} \bibinfo{year}{2011}\natexlab{}.
\newblock \showarticletitle{Quantifying location privacy}. In
  \bibinfo{booktitle}{\emph{2011 IEEE symposium on security and privacy}}.
  IEEE, \bibinfo{pages}{247--262}.
\newblock


\bibitem[Sweeney(2002a)]%
        {sweeney2002achieving}
\bibfield{author}{\bibinfo{person}{Latanya Sweeney}.}
  \bibinfo{year}{2002}\natexlab{a}.
\newblock \showarticletitle{Achieving k-anonymity privacy protection using
  generalization and suppression}.
\newblock \bibinfo{journal}{\emph{International Journal of Uncertainty,
  Fuzziness and Knowledge-Based Systems}} \bibinfo{volume}{10},
  \bibinfo{number}{05} (\bibinfo{year}{2002}), \bibinfo{pages}{571--588}.
\newblock


\bibitem[Sweeney(2002b)]%
        {sweeney2002k}
\bibfield{author}{\bibinfo{person}{Latanya Sweeney}.}
  \bibinfo{year}{2002}\natexlab{b}.
\newblock \showarticletitle{k-anonymity: A model for protecting privacy}.
\newblock \bibinfo{journal}{\emph{International journal of uncertainty,
  fuzziness and knowledge-based systems}} \bibinfo{volume}{10},
  \bibinfo{number}{05} (\bibinfo{year}{2002}), \bibinfo{pages}{557--570}.
\newblock


\bibitem[Talukder and Ahamed(2010)]%
        {talukder2010preventing}
\bibfield{author}{\bibinfo{person}{Nilothpal Talukder} {and}
  \bibinfo{person}{Sheikh~Iqbal Ahamed}.} \bibinfo{year}{2010}\natexlab{}.
\newblock \showarticletitle{Preventing multi-query attack in location-based
  services}. In \bibinfo{booktitle}{\emph{Proceedings of the third ACM
  conference on Wireless network security}}. \bibinfo{pages}{25--36}.
\newblock


\bibitem[Tamersoy et~al\mbox{.}(2012)]%
        {tamersoy2012anonymization}
\bibfield{author}{\bibinfo{person}{Acar Tamersoy}, \bibinfo{person}{Grigorios
  Loukides}, \bibinfo{person}{Mehmet~Ercan Nergiz}, \bibinfo{person}{Yucel
  Saygin}, {and} \bibinfo{person}{Bradley Malin}.}
  \bibinfo{year}{2012}\natexlab{}.
\newblock \showarticletitle{Anonymization of longitudinal electronic medical
  records}.
\newblock \bibinfo{journal}{\emph{IEEE Transactions on Information Technology
  in Biomedicine}} \bibinfo{volume}{16}, \bibinfo{number}{3}
  (\bibinfo{year}{2012}), \bibinfo{pages}{413--423}.
\newblock


\bibitem[Tu et~al\mbox{.}(2017)]%
        {tu2017beyond}
\bibfield{author}{\bibinfo{person}{Zhen Tu}, \bibinfo{person}{Kai Zhao},
  \bibinfo{person}{Fengli Xu}, \bibinfo{person}{Yong Li}, \bibinfo{person}{Li
  Su}, {and} \bibinfo{person}{Depeng Jin}.} \bibinfo{year}{2017}\natexlab{}.
\newblock \showarticletitle{Beyond k-anonymity: protect your trajectory from
  semantic attack}. In \bibinfo{booktitle}{\emph{2017 14th Annual IEEE
  International Conference on Sensing, Communication, and Networking
  (SECON)"}}. IEEE, \bibinfo{pages}{1--9}.
\newblock


\bibitem[Tu et~al\mbox{.}(2019)]%
        {8506438}
\bibfield{author}{\bibinfo{person}{Zhen Tu}, \bibinfo{person}{Kai Zhao},
  \bibinfo{person}{Fengli Xu}, \bibinfo{person}{Yong Li}, \bibinfo{person}{Li
  Su}, {and} \bibinfo{person}{Depeng Jin}.} \bibinfo{year}{2019}\natexlab{}.
\newblock \showarticletitle{Protecting Trajectory From Semantic Attack
  Considering ${k}$ -Anonymity, ${l}$ -Diversity, and ${t}$ -Closeness}.
\newblock \bibinfo{journal}{\emph{IEEE Transactions on Network and Service
  Management}} \bibinfo{volume}{16}, \bibinfo{number}{1}
  (\bibinfo{year}{2019}), \bibinfo{pages}{264--278}.
\newblock
\urldef\tempurl%
\url{https://doi.org/10.1109/TNSM.2018.2877790}
\showDOI{\tempurl}


\bibitem[Wernke et~al\mbox{.}(2014)]%
        {wernke2014classification}
\bibfield{author}{\bibinfo{person}{Marius Wernke}, \bibinfo{person}{Pavel
  Skvortsov}, \bibinfo{person}{Frank D{\"u}rr}, {and} \bibinfo{person}{Kurt
  Rothermel}.} \bibinfo{year}{2014}\natexlab{}.
\newblock \showarticletitle{A classification of location privacy attacks and
  approaches}.
\newblock \bibinfo{journal}{\emph{Personal and ubiquitous computing}}
  \bibinfo{volume}{18} (\bibinfo{year}{2014}), \bibinfo{pages}{163--175}.
\newblock


\bibitem[Xu and Cai(2008)]%
        {xu2008exploring}
\bibfield{author}{\bibinfo{person}{Toby Xu} {and} \bibinfo{person}{Ying Cai}.}
  \bibinfo{year}{2008}\natexlab{}.
\newblock \showarticletitle{Exploring historical location data for anonymity
  preservation in location-based services}. In \bibinfo{booktitle}{\emph{IEEE
  INFOCOM 2008-The 27th Conference on Computer Communications}}. IEEE,
  \bibinfo{pages}{547--555}.
\newblock


\bibitem[Yaseen et~al\mbox{.}(2018)]%
        {yaseen2018improved}
\bibfield{author}{\bibinfo{person}{Saba Yaseen}, \bibinfo{person}{Syed M~Ali
  Abbas}, \bibinfo{person}{Adeel Anjum}, \bibinfo{person}{Tanzila Saba},
  \bibinfo{person}{Abid Khan}, \bibinfo{person}{Saif Ur~Rehman Malik},
  \bibinfo{person}{Naveed Ahmad}, \bibinfo{person}{Basit Shahzad}, {and}
  \bibinfo{person}{Ali~Kashif Bashir}.} \bibinfo{year}{2018}\natexlab{}.
\newblock \showarticletitle{Improved generalization for secure data
  publishing}.
\newblock \bibinfo{journal}{\emph{IEEE Access}}  \bibinfo{volume}{6}
  (\bibinfo{year}{2018}), \bibinfo{pages}{27156--27165}.
\newblock


\bibitem[Yuan et~al\mbox{.}(2011)]%
        {yuan2011driving}
\bibfield{author}{\bibinfo{person}{Jing Yuan}, \bibinfo{person}{Yu Zheng},
  \bibinfo{person}{Xing Xie}, {and} \bibinfo{person}{Guangzhong Sun}.}
  \bibinfo{year}{2011}\natexlab{}.
\newblock \showarticletitle{Driving with knowledge from the physical world}. In
  \bibinfo{booktitle}{\emph{Proceedings of the 17th ACM SIGKDD international
  conference on Knowledge discovery and data mining}}.
  \bibinfo{pages}{316--324}.
\newblock


\bibitem[Yuan et~al\mbox{.}(2010)]%
        {yuan2010t}
\bibfield{author}{\bibinfo{person}{Jing Yuan}, \bibinfo{person}{Yu Zheng},
  \bibinfo{person}{Chengyang Zhang}, \bibinfo{person}{Wenlei Xie},
  \bibinfo{person}{Xing Xie}, \bibinfo{person}{Guangzhong Sun}, {and}
  \bibinfo{person}{Yan Huang}.} \bibinfo{year}{2010}\natexlab{}.
\newblock \showarticletitle{T-drive: driving directions based on taxi
  trajectories}. In \bibinfo{booktitle}{\emph{Proceedings of the 18th
  SIGSPATIAL International conference on advances in geographic information
  systems}}. \bibinfo{pages}{99--108}.
\newblock


\bibitem[Zheng et~al\mbox{.}(2008)]%
        {zheng2008understanding}
\bibfield{author}{\bibinfo{person}{Yu Zheng}, \bibinfo{person}{Quannan Li},
  \bibinfo{person}{Yukun Chen}, \bibinfo{person}{Xing Xie}, {and}
  \bibinfo{person}{Wei-Ying Ma}.} \bibinfo{year}{2008}\natexlab{}.
\newblock \showarticletitle{Understanding mobility based on GPS data}. In
  \bibinfo{booktitle}{\emph{Proceedings of the 10th international conference on
  Ubiquitous computing}}. \bibinfo{pages}{312--321}.
\newblock


\bibitem[Zheng et~al\mbox{.}(2010)]%
        {zheng2010geolife}
\bibfield{author}{\bibinfo{person}{Yu Zheng}, \bibinfo{person}{Xing Xie},
  \bibinfo{person}{Wei-Ying Ma}, {et~al\mbox{.}}}
  \bibinfo{year}{2010}\natexlab{}.
\newblock \showarticletitle{GeoLife: A collaborative social networking service
  among user, location and trajectory.}
\newblock \bibinfo{journal}{\emph{IEEE Data Eng. Bull.}} \bibinfo{volume}{33},
  \bibinfo{number}{2} (\bibinfo{year}{2010}), \bibinfo{pages}{32--39}.
\newblock


\bibitem[Zheng et~al\mbox{.}(2009)]%
        {zheng2009mining}
\bibfield{author}{\bibinfo{person}{Yu Zheng}, \bibinfo{person}{Lizhu Zhang},
  \bibinfo{person}{Xing Xie}, {and} \bibinfo{person}{Wei-Ying Ma}.}
  \bibinfo{year}{2009}\natexlab{}.
\newblock \showarticletitle{Mining interesting locations and travel sequences
  from GPS trajectories}. In \bibinfo{booktitle}{\emph{Proceedings of the 18th
  international conference on World wide web}}. \bibinfo{pages}{791--800}.
\newblock


\end{thebibliography}

\end{document}